%% file: paper_arxiv.tex
\documentclass[10pt,a4paper]{article}
\usepackage[utf8]{inputenc}
\usepackage{amsmath}
\usepackage{amsfonts}
\usepackage{amssymb}

\usepackage{graphicx}
\usepackage{caption}
\usepackage{subcaption}

\usepackage{epstopdf}
\usepackage{abstract}
\usepackage[toc,page]{appendix}
\usepackage[sort]{cite}
\usepackage{authblk}
\usepackage{setspace}
\usepackage{color}

\usepackage[numbers,super,compress]{natbib}
\usepackage{longtable}

\usepackage{hyperref}

\newcommand{\bvec}[1]{\mathbf{#1}}

\newcommand{\colornote}[0]{(Colour online.) }
\newcommand{\eg}[0]{\emph{e.g.}}

\graphicspath{{./images/}}

\title{ Energetically favoured defects in dense packings of particles on spherical surfaces } 
\author[1]{Stefan Paquay}
\author[2]{Halim Kusumaatmaja}
\author[3]{David J. Wales}
\author[4]{Roya Zandi}
\author[1,5]{Paul van der Schoot}
\affil[1]{Department of Applied Physics, Technische Universiteit Eindhoven, The Netherlands}
\affil[2]{Department of Physics, University of Durham, UK}
\affil[3]{Department of Chemistry, University of Cambridge, UK}
\affil[4]{Department of Physics and Astronomy, University of California, Riverside, USA}
\affil[5]{Instituut voor Theoretische Fysica, Universiteit Utrecht, The Netherlands}
\date{  }

\begin{document}
  \maketitle
  \begin{abstract}
    The dense packing of interacting particles on spheres has proved to be a useful model for virus capsids and colloidosomes.
    Indeed, icosahedral symmetry observed in virus capsids corresponds to potential energy minima that occur for magic numbers of, \emph{e.g.}, 12, 32 and 72 identical Lennard-Jones particles,
for which the packing has exactly the minimum number of twelve five-fold defects.
    It is unclear, however, how stable these structures are against thermal agitation.
    We investigate this property by means of basin-hopping global optimisation and Langevin dynamics for particle numbers between ten and one hundred. An important measure is the number and type of point defects, that is, particles that do not have six nearest neighbours.
    We find that small icosahedral structures are the most robust against thermal fluctuations, exhibiting fewer excess defects and rearrangements for a wide temperature range.
    Furthermore, we provide evidence that excess defects appearing at low non-zero temperatures lower the potential energy at the expense of entropy. At higher temperatures defects are, as expected, thermally excited and thus entropically stabilised.
    If we replace the Lennard-Jones potential by a very short-ranged (Morse) potential, which is arguably more appropriate for colloids and virus capsid proteins, we find that the same particle numbers give a minimum in the potential energy, although for larger particle numbers these minima correspond to different packings.
    Furthermore, defects are more difficult to excite thermally for the short-ranged potential, suggesting that the short-ranged interaction further stabilises equilibrium structures.
  \end{abstract}

\section{Introduction}

Virus capsids \cite{caspar-1962} and colloidosomes \cite{dinsmore-2002} have been succesfully modelled as dense packings of spherical particles constrained to a spherical surface, in particle-based \cite{bruinsma-2004,zandi-2004,fantoni-2012} and phase-field calculations.\cite{backofen-2010} The equilibrium packings follow from the interplay between the curvature of the sphere and the interaction between the particles. For fixed particle size and surface coverage, increasing the radius of curvature of the surface leads to packings that exhibit varying numbers of isolated point defects that for large enough particle numbers condense into clusters of defects.\cite{bausch-2003,backofen-2010} Here, defects are particles that do not have the ideal six-fold coordination.
Studies of particles on unduloids and catenoids have shown that for small particle numbers a Lennard-Jones potential produces different minimum energy structures compared to a purely repulsive Coulomb potential, showing that the range and type of interaction also affect the geometry of particle packings on curved surfaces.\cite{kusumaatmaja-2013} For packings on spherical surfaces, the minimum energy structures for 12, 24, 32, 44 and 48 particles are the same for the Lennard-Jones and repulsive Coulomb potential, whereas for many others, including 72, these are different.\cite{voogd-thesis,wales-2006}


In their study of why spherical viruses almost invariably exhibit icosahedral symmetry, Zandi \emph{et al.} \cite{zandi-2004} found by Monte Carlo simulation of Lennard-Jones particles on a spherical surface that, if the particle number allows it, the equilibrium packings do in fact have icosahedral symmetry. This effect occurs for the magic particle numbers $N=12,~32$ and $72,$ corresponding to $T=1,~3$ and 7 icosahedral symmetry. By allowing a switch between larger and smaller particle sizes, modeling pentameric and hexameric capsomeres, icosahedral symmetry is also recovered for $N=42,$ which is the $T=4$ structure.
Fejer \emph{et al.} studied a different model of rigid bodies consisting of an attractive disk and two repulsive Lennard-Jones axial sites on top and bottom. These sites induce a preferred curvature. In this model, icosahedral packings turn out to be local potential energy minima for $N=12, 32$ and 72, but they found that the $T=4$ icosahedral symmetry for $N=42$ is only a minimum energy structure if the disks assemble on top of a template.\cite{fejer-2010}
In the single-particle description that we follow, all other particle numbers give non-icosahedral structures, often with more than the minimum required twelve five-fold point defects. 

Apparently, even for a single particle size, the icosahedrally packed structures have a lower potential energy per particle than the packings of adjacent particle numbers, at least for the low non-zero temperatures considered.\cite{zandi-2004} This result suggests that viruses prefer icosahedral symmetry simply because this is the most optimal packing for the effective interaction between the capsomeres. The Monte Carlo simulations of Ref. \cite{zandi-2004} are consistent with the zero-temperature simulated annealing studies of Lennard-Jones particles packings by Voogd,\cite{voogd-thesis} in the sense that they recover potential energy minima at the same particle numbers. However, the latter study provides more detail about the symmetry of all the particle packings found. Interestingly, Voogd identifies the global minimum for the $N=72$ packing with a $D_{5h}$ point group, rather than an icosahedral one, which is one of the structures that Zandi \emph{et al.} identified at this size. This discrepancy could be due to the non-zero temperature in the simulations of Zandi \emph{et al.}, hinting at the potential importance of entropy. Indeed, our calculations of the potential energy for both packings confirm that the $D_{5h}$ packing has lower potential energy than the icosahedral one.


This analysis suggests that temperature could play an important role in the thermodynamic stability of the symmetry of dense packings of particles on a spherical surface. For non-zero temperature, minimum energy does not imply minimum free energy. Indeed, our computer ``experiments'' reveal that for certain numbers of Lennard-Jones particles confined to a spherical surface, energy favours excess defects, \emph{i.e.}, these packings have more than twelve defects for very low temperatures. Such energetically stabilised defects also appear for the Thomson problem \cite{wales-2006,wales-2009} and as grain boundary scars.\cite{bausch-2003}
Of course, at higher temperatures, entropy favours excess defects, in the form of thermally excited dislocations and/or disclinations analogous to melting in a 2D flat surface. For an extensive discussion we refer to the review of Strandburg.\cite{strandburg-1988}

Another question that arises is how representative the atomic Lennard-Jones potential is for interactions between complex particles such as proteins and colloids, and how sensitive the structure of dense particle packings on curved templates is to the shape of the potential. This question is relevant because interactions between proteins are arguably better described by a short-ranged potential, \cite{lomakin-2003,kegel-2004,prinsen-2006} and Van der Waals interactions between colloids are also shorter-ranged (stickier) than predicted by a Lennard-Jones potential.\cite{parsegian-boek} For example, the colloidosomes of the Manoharan group are induced by the presence of polymer molecules that give rise to extremely short-ranged depletion interactions between the colloids.\cite{meng-2014} For three-dimensional clusters it is already known that the range of the potential strongly influences the potential energy landscape. Doye \emph{et al.} have shown that the shorter ranged the attractive part of the potential, the larger the number of local energy minima for a given number of particles.\cite{doye-1995} 
 Furthermore, for small clusters of short-ranged particles it was found that temperature has a significant influence on the relative stability of different packings.\cite{wales-2010}

To address this issue in the context of particles confined to spherical surfaces, we consider a Morse potential of much shorter range than the Lennard-Jones form. For particle numbers of 32, and 24 and below, we find for the same particle numbers deep local potential energy minima that also turn out to have the same structure.
For larger particle numbers, the Morse potential produces deeper local minima in the potential energy landscape as function of the particle number. Furthermore, for those particle numbers that are a local minimum in the potential energy for both the Morse and Lennard-Jones potential, the particle arrangement proved different. Hence, for a shorter-ranged potential, for the same particle numbers, different packings minimise the potential energy. For Morse particles it also proved more difficult to thermally excite defects, indicating that a shorter-ranged potential stabilises the structures. This property is especially clear in the case of the $T=3$ icosahedron for $N=32.$ 

However, we find that the $T=7$ icosahedron for $N=72$ particles is no longer an equilibrium packing, nor a potential energy minimum. Thus, while the range of the interaction potential broadens the temperature range over which structures are stable, it also influences the symmetry of the equilibrium packing itself.
A similar observation was reported for simulations of disks with an adhesive edge confined to a spherical surface were performed.\cite{bruinsma-2004} For adhesive disks, the effective range of attraction is zero, and in this case, both $N=32$ and $N=72$ lose icosahedral symmetry. Thus, although a shorter range appears to help stabilise the equilibrium structures over a larger temperature range, it also changes the symmetry of the equilibrium packing. 


Perhaps our most counterintuitive result is that upon \emph{reducing} the temperature, excess defects appear for certain particle numbers.
In other words, for these particular sizes, the number of defects is a non-monotonic function of the temperature. This result must mean that with decreasing temperature these defects lower the total potential energy at the expense of entropy.
We find that this is also the case for $N=72,$ explaining the structural fluctuations observed by Zandi \emph{et al.},\cite{zandi-2004} which are due to the transition from an icosahedral symmetry to a $D_{5h}$ symmetry, and, as we shall see below, a $D_3$ symmetry. The $D_{5h}$ structure is the global energy minimum at zero temperature according to Voogd \cite{voogd-thesis} and according to our own basin-hopping calculations.

Because in our simulations the particles fluctuate between different packings, we can obtain free energy differences simply by determining the probability of finding each packing. From this probability we determine that the icosahedral packing, which has the fewest defects, is indeed entropically more favourable than the $D_{5h},$ confirming that the ground state can exhibit excess defects, similar to experimental observations and computational results for very much larger systems in the form of grain boundary scars \cite{bausch-2003,backofen-2010} and for packings of electrons on a sphere (the Thomson problem).\cite{wales-2006,wales-2009}

The remainder of this paper is laid out as follows. First we describe in Section \ref{sec:methods} the computational methods we employed. We also provide a discussion of how we quantify defects and how we determine them. Then, in Section \ref{sec:defects-lj} we discuss how temperature influences the stability of packings of Lennard-Jones particles. In Section \ref{sec:lj-free-energy} we discuss the appearance of defects in the ground state and determine free energy differences between the packings based on how often they are encountered. We continue in Section \ref{sec:defects-morse} to show that the equilibrium structures of Morse particles are much more robust against thermal fluctuations than those of Lennard-Jones particles, but that the minimum energy packings tend to differ from the Lennard-Jones packings at larger particle numbers. Finally, in Section \ref{sec:conclusion}, we underline the most important implications of the three different aspects of this work discussed above.

\section{Methods}
\label{sec:methods}
We consider packings of two different types of particle on a spherical surface. The first model employs the well-known Lennard-Jones potential, allowing us to directly compare our results with those of Zandi \emph{et al.} \cite{zandi-2004} and Voogd.\cite{voogd-thesis} We write the Lennard-Jones potential in terms of the equilibrium spacing, $r_0,$ rather than the more usual zero-potential distance, to allow for a straightforward comparison with the Morse potential later on. Specifically, we have
\begin{equation}
  V_{LJ}(r) = \epsilon \left[ \left(\frac{r_0}{r} \right)^{12} - 2\left(\frac{r_0}{r} \right)^{6}\right]. \label{eqn:lj}
\end{equation}
The potential has its most negative value $-\epsilon$ at $r=r_0,$ so $\epsilon$ can be treated as the interaction strength or pair well depth.
The second model employs the Morse potential
\begin{equation}
  V_M(r) = \epsilon \left[ e^{-2\alpha(r-r_0)} - 2 e^{-\alpha(r-r_0)}\right].
  \label{eqn:morse}
\end{equation}
In Eq. \eqref{eqn:morse}, the parameters $\epsilon$ and $r_0$ have the same meaning as in Eq. \eqref{eqn:lj}, but now there is an additional parameter $\alpha,$ which can be used to tune the interaction range. In this work we set $\alpha$ to a specific value to model the interaction potential induced by depletants that the Manoharan group put forward to discuss their experiments on colloidosomes.\cite{meng-2014}
We do this by fixing the ratio of the distances at which the potential has its most negative value and where it is only one tenth of that well depth. Applying this procedure leads to a value for the range parameter of $\alpha = (61.2\pm2)/r_0,$ which for convenience we rounded to $\alpha = 60/r_0.$ Such a large value for $\alpha$ leads to a much faster decay in the interaction strength and destabilises the liquid phase.\cite{hagen-1993,doye-1996,doye-1996-2,meng-2014}

In our Langevin dynamics simulations, performed with the LAMMPS program,\cite{plimpton-1995} we truncate and shift the potential at some cut-off distance $r_c$ by defining as actual interaction potential $V(r) = V_{LJ/M} (r) - V_{LJ/M}(r_c),$ where the subscript $LJ$ denotes the Lennard-Jones potential and subscript $M$ the Morse potential. We take as time unit the Langevin damping time $\tau_L,$ which describes the time over which the velocity autocorrelation decays. For our purposes, the exact value of the damping time should be irrelevant because all our simulations focus on systems under conditions of thermodynamic equilibrium. We take $r_c = 2.5 r_0/2^{1/6} \approx 2.2272 r_0,$ at which the untruncated Lennard-Jones and Morse potentials have values of $-0.016~\epsilon$ and $-2.1~10^{-32}~\epsilon$ respectively. The distance $r_c$ corresponds to a cut-off at exactly $2.5\sigma$ in terms of the more common Lennard-Jones distance parameter $\sigma.$

Care was taken to ensure that the centre of mass of all particles does not acquire an angular momentum from coupling to the thermostat. This restriction is achieved by subtracting in each step from all particle velocities, the vector $\boldsymbol{\omega}_{CM} \times \bvec{x}_i/N$ with $\boldsymbol{\omega}_{CM}$ the angular velocity of the centre of mass, $\bvec{x}_i$ the position vector of particle $i,$ and $N$ the number of particles. After subtracting this component, the velocities are all rescaled such that the kinetic energy before and after the correction is unchanged. Note, however, that the kinetic energy is not constant, as the Langevin thermostat imposes fluctuations consistent with the canonical ensemble. Because the particles are constrained to a sphere, there is no need to subtract the linear velocity of the centre of mass.

For both potentials, we attempt to find for all $N=10$ to $N=100$ the global potential energy minimum using the basin-hopping method \cite{wales-1997} as well as thermal equilibrium packings in a temperature range between $T=0.001~\epsilon/k_B $ and $T = 2~\epsilon/k_B,$ where $k_B$ is Boltzmann's constant. For each $N$ a surface density $\rho$ has to be chosen. Let $R$ be the radius of the spherical surface. Then $\rho = N / 4 \pi R^2,$ and $R$ has to be determined for each $N.$ A natural choice for $R$ is the radius that results in the lowest potential energy at zero temperature. For Lennard-Jones potentials, these radii are tabulated by Voogd in \cite{voogd-thesis} and are consistent with our basin-hopping calculations, but for a Morse potential we have not been able to find tabulated values. We therefore employ the following strategy.

We perform Langevin dynamics simulations of $N$ particles constrained to a sphere using a special case of the RATTLE algorithm \cite{andersen-1983} described in,\cite{paquay-2016} in which we linearly shrink the radius from an initial value $R_0$ to a final value $R_1$ over a time span equal to $10^4\tau_{L}.$ The values for $R_0$ and $R_1$ we estimate from considerations on hard disk packings, which gives rise to a natural sphere radius $R^*.$ To calculate $R^*,$ consider $N$ hard disks of diameter $d_0$ that cover an area fraction $\phi = N d_0^2/ 16R^2$ of the sphere. The upper limit to $\phi$ in a flat, two-dimensional geometry is equal to $\phi_m = \pi / \sqrt{12}.$\cite{bruinsma-2004} The radius that gives this maximum is then $R^* = d_0\sqrt{N/\phi_m}/4.$ With $d_0$ we associate the minimum of the interaction potential $r_0,$ because for $r<r_0$ both potentials are steeply repulsive.

We search for a minimum in the potential energy around $R^*$ by putting $R_0 = 1.3R^*$ and $R_1 = 0.8R^*.$ For each $N$ we monitor over time the potential energy and radius of the spherical template as it shrinks from $R_1$ to $R_0.$ This schedule produces an energy trace for each $N$ as a function of $R$ similar to those presented by Voogd,\cite{voogd-thesis} with a characteristic deep minimum just before a steep increase for small $R,$ from which the optimal radius can be determined with a simple post-processing script.

\begin{figure}[htb]
  \centering
  \includegraphics[width=0.8\textwidth]{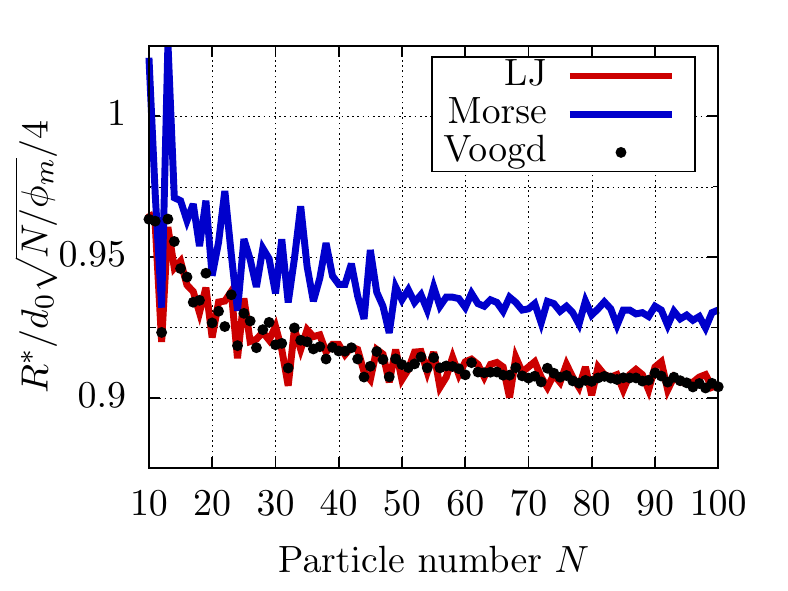}
  \caption{\colornote{}Sphere radii $R^*$ that minimise the potential energy for $N$ particles interacting either through a Lennard-Jones (LJ) or a Morse (Morse) potential, as fraction of the estimated radius that would tightly pack $N$ hard disks of diameter $d_0$, $d_0\sqrt{N/\phi_M}/4$. Note that the LJ data coincides well with the results of Voogd.\cite{voogd-thesis} The largest difference between the two is no more than $0.02r_0$ ($<2\%$). \label{fig:min-radii}}
\end{figure}

We present the optimal radii $R$ as function of $N$ in Fig. \ref{fig:min-radii} for both the Lennard-Jones particles and the Morse particles for the case $\alpha = 60/r_0.$ Note that for the Morse particles, the sphere radius is larger for all $10 \leq N \leq 100,$ because the penalty for overlap is much larger and cannot be compensated easily by next-nearest neighbour interactions. For the Lennard-Jones particles, the difference in the optimal radius $R$ between our data and Voogd's is less than 2\% for all $N$ and the largest deviation in total energy is below 1\%. Additionally, the potential energies we find at the optimal radius match closely with those presented by Zandi \emph{et al.} in.\cite{zandi-2004} Furthermore, if we use the same method of quantifying defects as Voogd,\cite{voogd-thesis} which is based on Voronoi constructions, we find the same distribution of topological charges, reassuring us that we obtain the same structures. For a complete tabulation of our energies and sphere radii, see SI 1.

We also consider another way to quantify defects.
In a 3D Voronoi construction, the entire simulation volume $V$ is divided into $N$ polyhedra with volume $V_i,$ one for each particle $i=1,...,N$ (a Voronoi tesselation). Each volume $V_i$ consists of all points $\bvec{x}$ that are closer to the position of particle $\bvec{x}_i$ than to any other particle. Although our particles only have access to a two-dimensional subspace of $\mathbb{R}^3,$ they live in a three-dimensional Cartesian space, so it is still possible to assign the aforementioned volumes to them. The number of faces of each polyhedron is then the number of nearest neighbours of each particle, and a connectivity network can be generated by connecting all particles whose polyhedra share a face.

Voronoi analysis is a natural way to determine nearest neighbours in a hexagonal lattice, as the polyhedra in a 2D hexagonal lattice will be hexagons as well, even when there are significant thermal fluctuations in the particle positions. The network generated by the tesselation covers the entire space, and thus automatically has the correct Euler characteristic. For particles on a sphere, the equivalent of the Voronoi tesselation can be obtained by determining the convex hull \cite{brown-1978} for which we employ the CGAL software library.\cite{cgal}

An issue arises with Voronoi tesselation when particles are packed in other types of lattice. For example, in a perfectly square lattice, the Voronoi tesselation is degenerate because the cubes around the particles have touching edges and vertices. A small thermal fluctuation will generate an additional face in two of the polyhedra, resulting in either one pair or the other being counted as neighbours, even though the packing should be considered square instead of hexagonal.\cite{troadec-1998} In a previous work on global energy minima of the Thomson problem such configurations were encountered,\cite{wales-2006,wales-2009} so this apparent pathology was anticipated in the present work.

In fact, we do encounter such structures for Lennard-Jones and Morse particles in our simulations.  Hence, we decided to also invoke a distance criterion strategy to determine the number of nearest neighbours. With this criterion, all particles that are within a certain distance $r^*$ of each other are considered nearest neighbours. In this case, care has to be taken to select a sensible value for $r^*.$ One way to do this is to determine for every $N$ at what distance the second minimum in the pair distribution function is located and to use that to fix $r^*.$ Some structures, however, produce a split first peak at around the minimum of the potential energy $r = r_0.$ In that case, we choose as $r^*$ a distance after the split peak but before the second major peak. 

In principle, $r^*$ is a function of temperature, so it should be determined for every temperature $T.$ For practical reasons, however, we determine $r^*$ only at the low temperature of $T = 0.01\epsilon/k_B.$ 
For $N=24,~32,~44,~48$ and 72 we verified that the $r^*$ obtained this way still coincides with a minimum in the time-averaged pair correlation function at a higher temperature of $T=0.5\epsilon/k_B.$
With the distance criterion, square lattices are identified more robustly in the presence of thermal fluctuations than by means of tesselation, especially at lower temperatures. See SI 2 for a more thorough description of this procedure and a tabulation of obtained cut-off radii $r^*.$ Note, however, that the network generated by connecting the nearest neighbours in general does not have the proper Euler characteristic, an issue we choose to ignore. Because of the drawbacks associated with both methods, we apply both and compare the results they provide.

Finally, for the representative case of $N=72$ particles, we determine the free energy difference between specific packings as a function of temperature to extract the relative contributions of potential energy and entropy. Our first attempts to determine these properties with thermodynamic integration as described in \cite{frenkel-boek} did not produce satisfactory results. However, since in our simulations the packings fluctuate between different symmetries, we count their occurrence frequencies. From these frequencies we can reconstruct at each temperature the probability of finding a packing. From the probability ratio for two different configurations, say, $a$ and $b$, we calculate a free energy difference. The probability $P_a$ of encountering $a$ scales with the Boltzmann factor as $P_a \sim \exp(-F_a/k_BT),$ where $F_a$ is the free energy of packing $a.$ Hence, the ratio of two of these probabilities is $P_a / P_b = \exp(-(F_a - F_b) / k_B T) = \exp( -\Delta F_{ab}/k_BT).$ In other words, $\Delta F_{ab} = -k_B T \ln(P_a/P_b).$ Entropy differences can be derived from the slope of the free energy difference as a function of temperature, since $S = -(\partial F/\partial T)_{N,R}$ evaluated at constant particle number $N$ and sphere radius $R.$

We next consider the thermal stability of Lennard-Jones packings in Section \ref{sec:defects-lj} by investigating the number of point defects at various temperatures. We then focus in Section \ref{sec:lj-free-energy} on some packings that have additional defects in their ground state, and we determine their stability at different temperatures by calculating their free energy. Finally we perform the same stability analysis for short-ranged Morse particles in Section \ref{sec:defects-morse}.

\section{Lennard-Jones defect landscape}
\label{sec:defects-lj}
We determine for our Lennard-Jones particles the excess number of point defects as a function of temperature and particle number. Excess point defects are particles that do not have six nearest neighbours in excess of the 12 that are required to satisfy the Euler criterion. We present results using both the distance criterion and the Voronoi triangulation. The data for $T=0$ are generated by means of basin-hopping calculations with the aid of the GMIN program.\cite{wales-1997,gmin} The data for $T>0$ are obtained from a Langevin dynamics simulation using the LAMMPS program.\cite{plimpton-1995} The damping time of the thermostat, $\tau_L,$ is the reference time unit, while the time step size is fixed at $0.005~\tau_{L}.$ We invoked the Gr\o{}nbech-Jensen-Farago formulation of the Langevin forces,\cite{gronbech-jensen-2013} which generates positions that are correctly Boltzmann-distributed for the thermostat temperature for larger time steps albeit at the expense of inaccuracies in the velocity distribution. Since none of the properties we are interested in depend on the velocity distribution, this is an acceptable drawback. 

\begin{figure}[tb]
  \centering
  \includegraphics[width=0.8\textwidth]{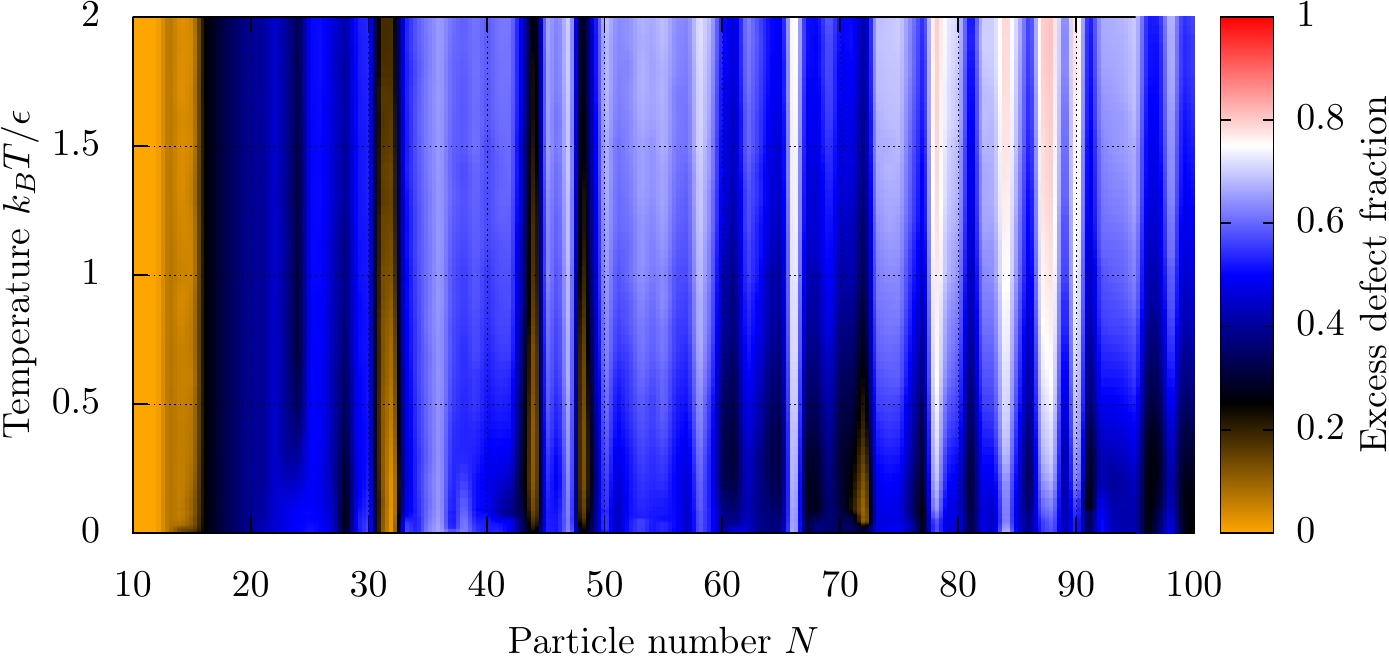}
  \caption{\colornote{}Temperature dependence of the excess defect fraction, the number of particles with other than six nearest neighbours beyond the first twelve, for $N=10$ to $N=100$ Lennard-Jones particles, using the distance criterion.  \label{fig:lj-dist-defects}}
\end{figure}

In Fig. \ref{fig:lj-dist-defects} we show the excess defect fraction, where we consider as neighbours all particles within the distance $r^*$ at which the pair distribution function has its second minimum. (See also Section \ref{sec:methods} for details and SI 2 for values of $r^*$ as a function of the particle number $N.$) The excess defect fraction is the number of point defects in excess of the first twelve, divided by the total number of particles. From Fig. \ref{fig:lj-dist-defects} we can see that for many particle numbers $N$ there are excess defects across the entire temperature range probed. It is also striking that $N=32$ is the only packing, apart from $N=12,$ that retains icosahedral symmetry. Indeed, for $N=32$ we find zero excess defects over a wide temperature range. The largest excess defect fraction is 0.2 for the highest temperature of $T = 2~\epsilon/k_B.$ This result indicates that the $T=3$ icosahedral structure of 32 particles is very robust against thermal fluctuations. 

\begin{figure}[htb]
  \centering
  \begin{subfigure}[t]{0.18\textwidth}
    \includegraphics[width=\textwidth]{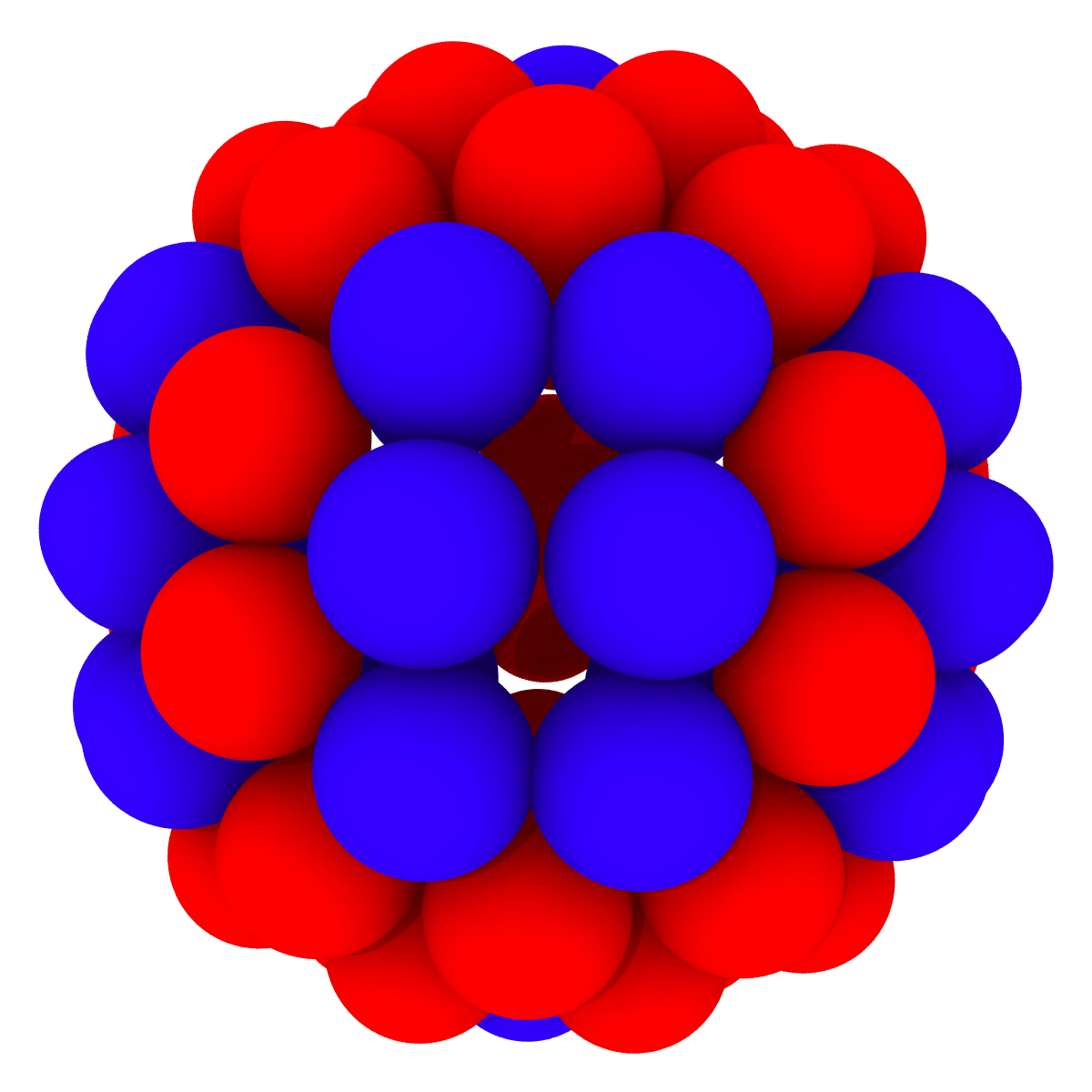}
  \end{subfigure}
  \begin{subfigure}[t]{0.18\textwidth}
    \includegraphics[width=\textwidth]{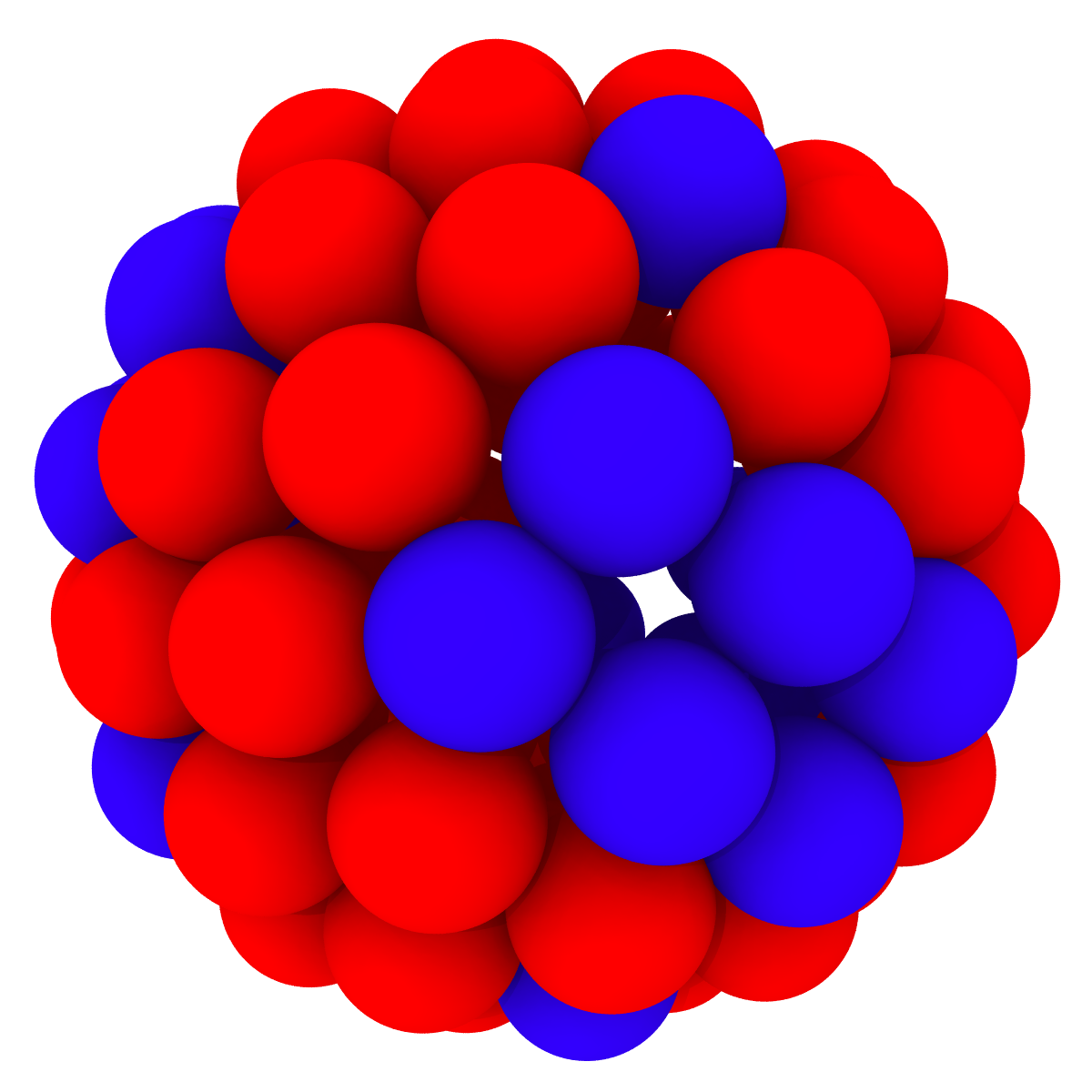}
  \end{subfigure}
  \begin{subfigure}[t]{0.18\textwidth}
    \includegraphics[width=\textwidth]{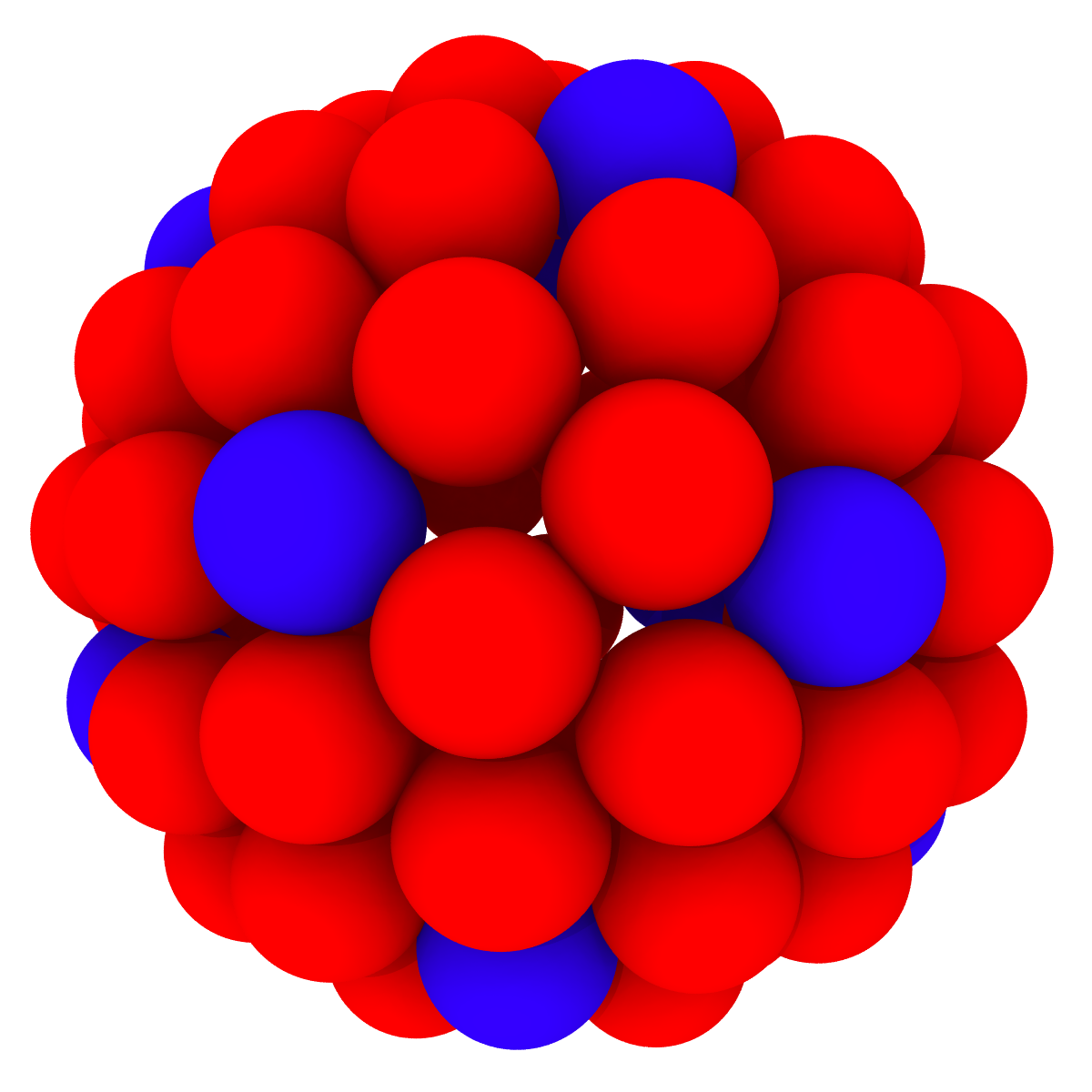}
  \end{subfigure}
  \begin{subfigure}[t]{0.18\textwidth}
    \includegraphics[width=\textwidth]{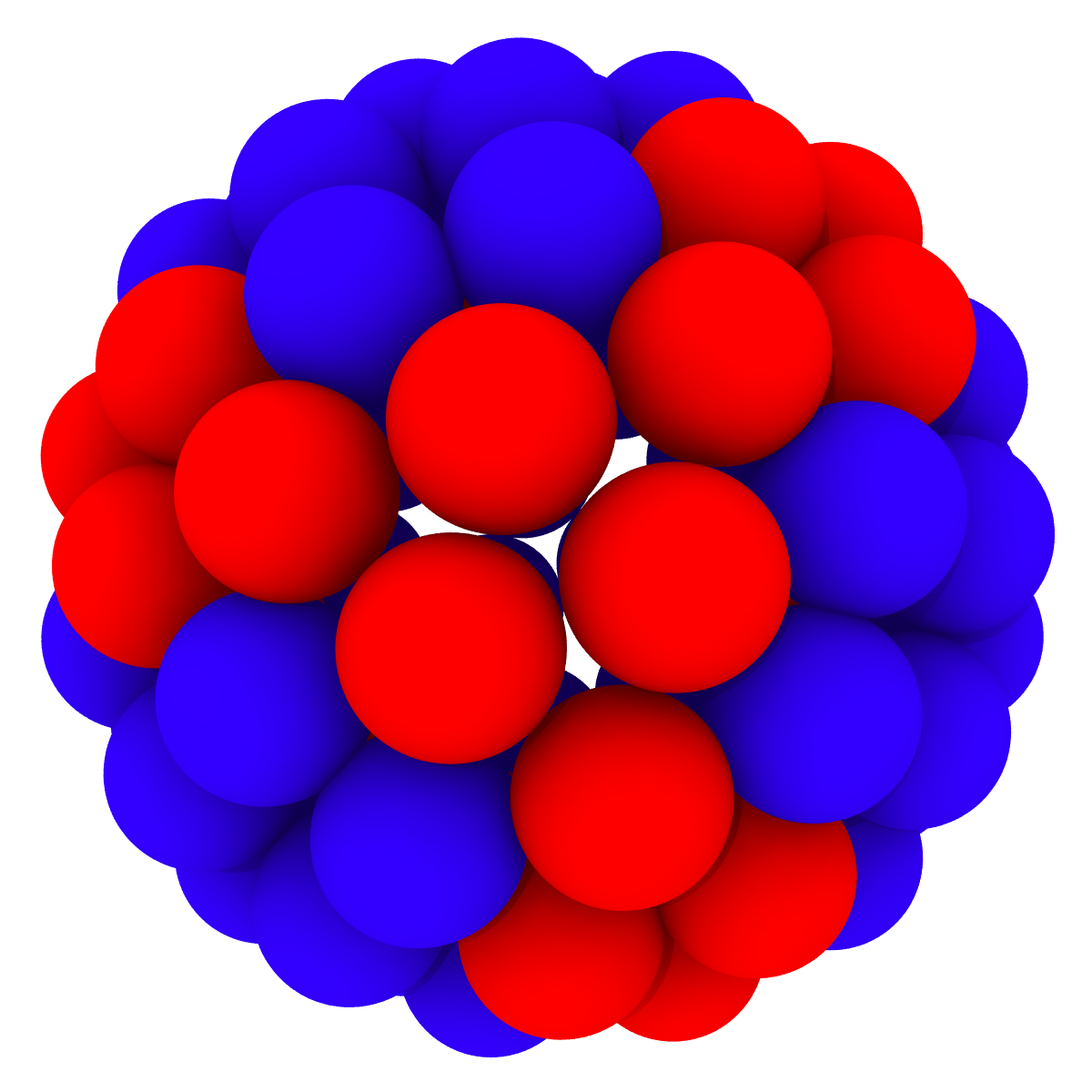}
  \end{subfigure}
  \begin{subfigure}[t]{0.18\textwidth}
    \includegraphics[width=\textwidth]{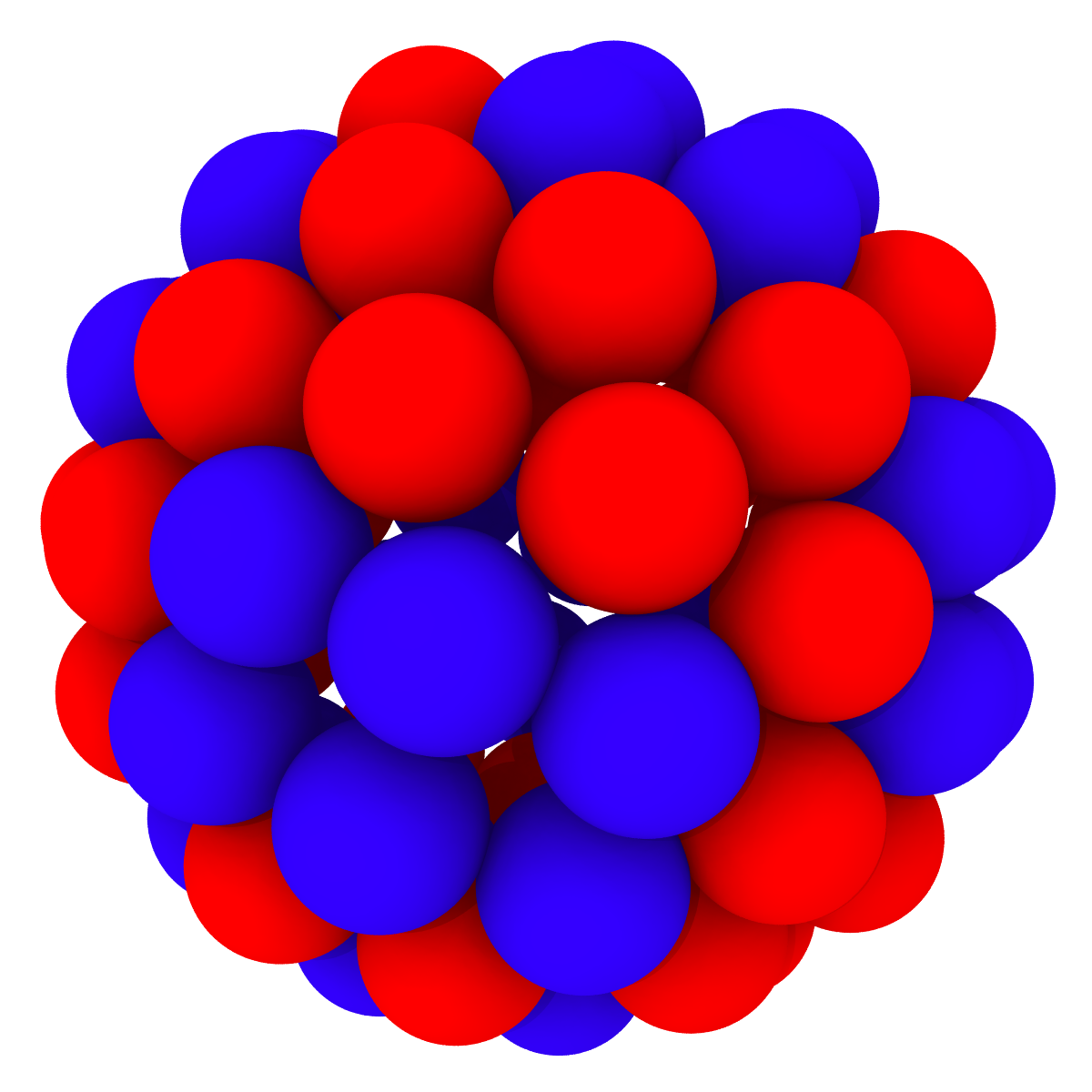}
  \end{subfigure} \\
  \begin{subfigure}[t]{0.18\textwidth}
    \includegraphics[width=\textwidth]{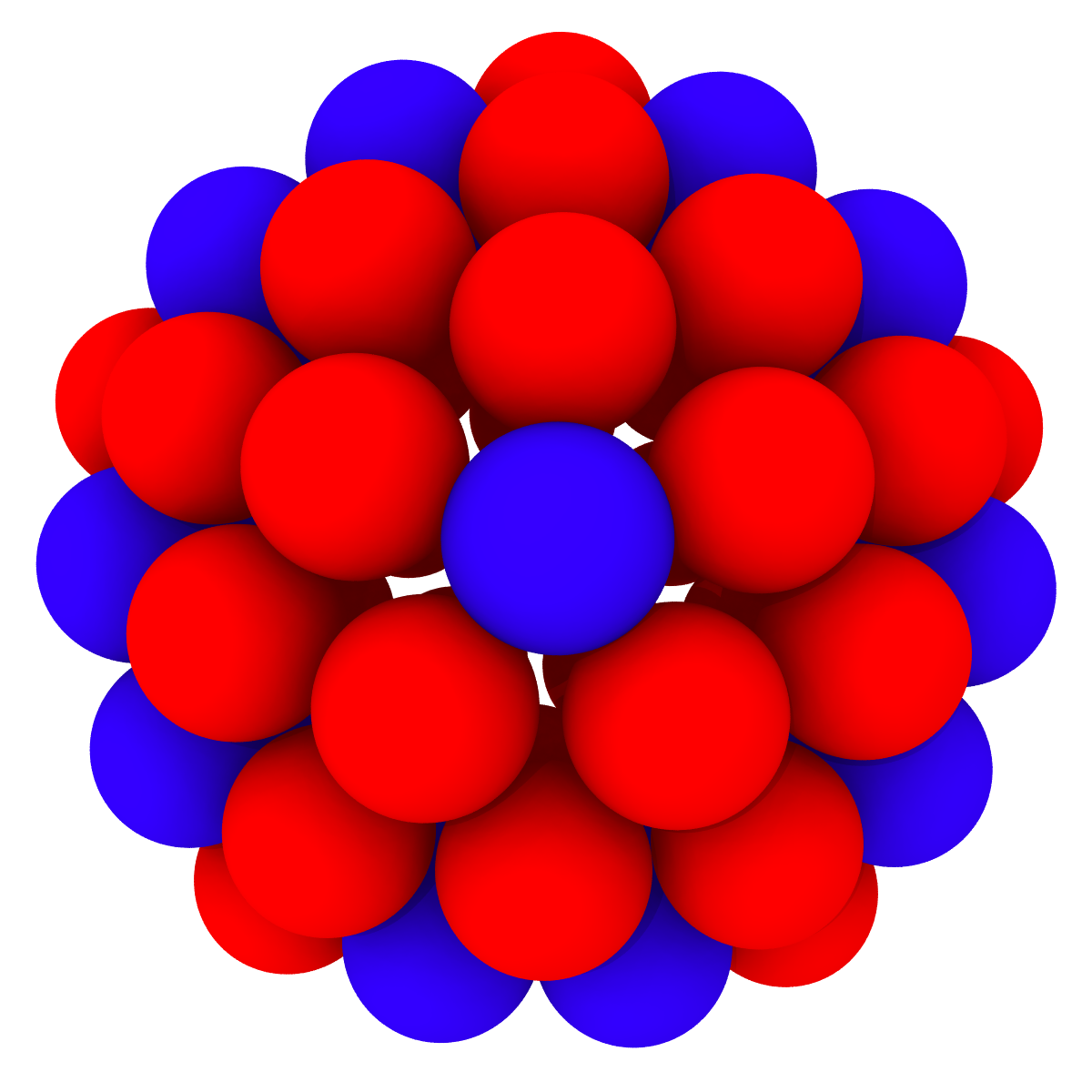}
    \caption{\label{fig:N72-min-D5h}}
  \end{subfigure}
  \begin{subfigure}[t]{0.18\textwidth}
    \includegraphics[width=\textwidth]{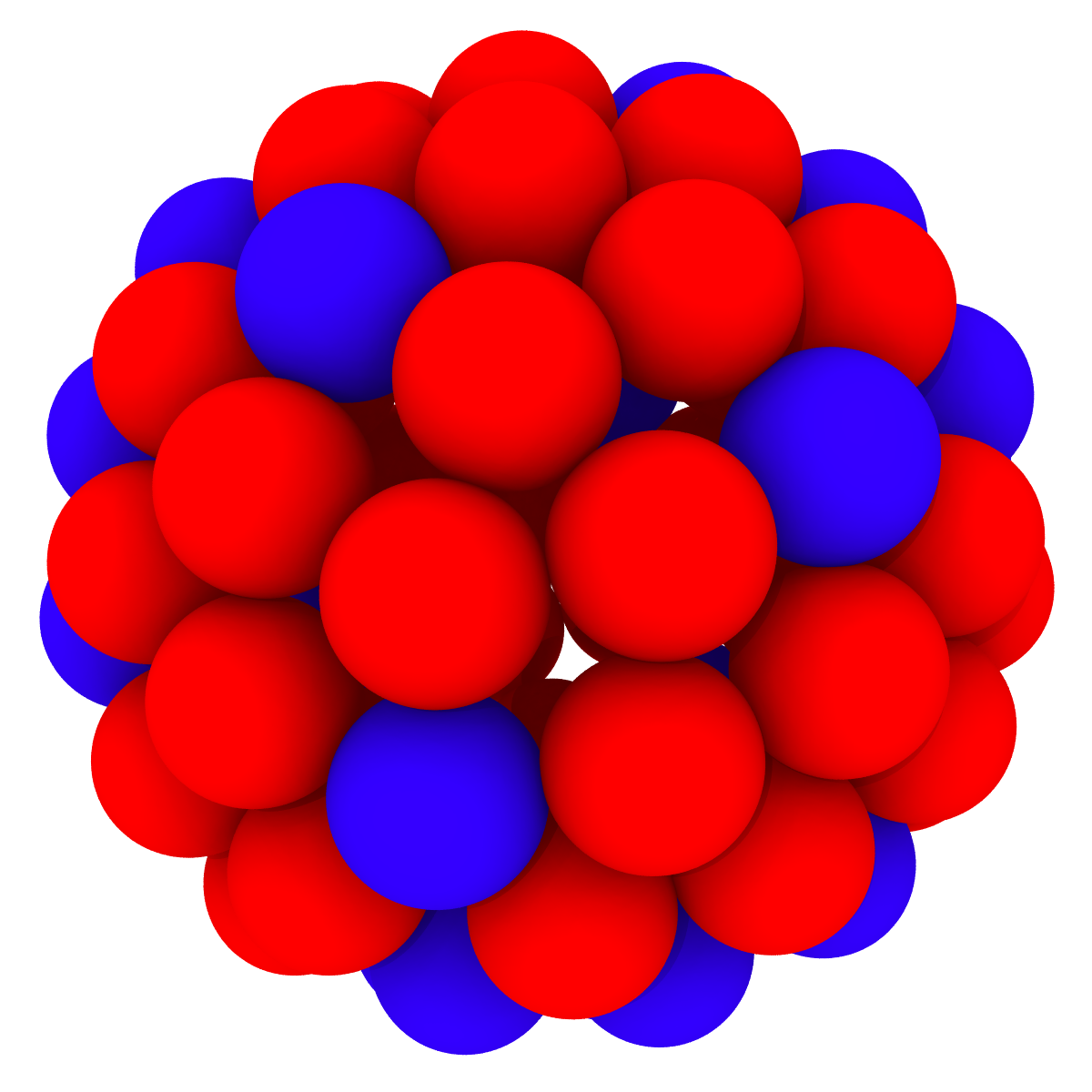}
    \caption{\label{fig:N72-min-D3}}
  \end{subfigure}
  \begin{subfigure}[t]{0.18\textwidth}
    \includegraphics[width=\textwidth]{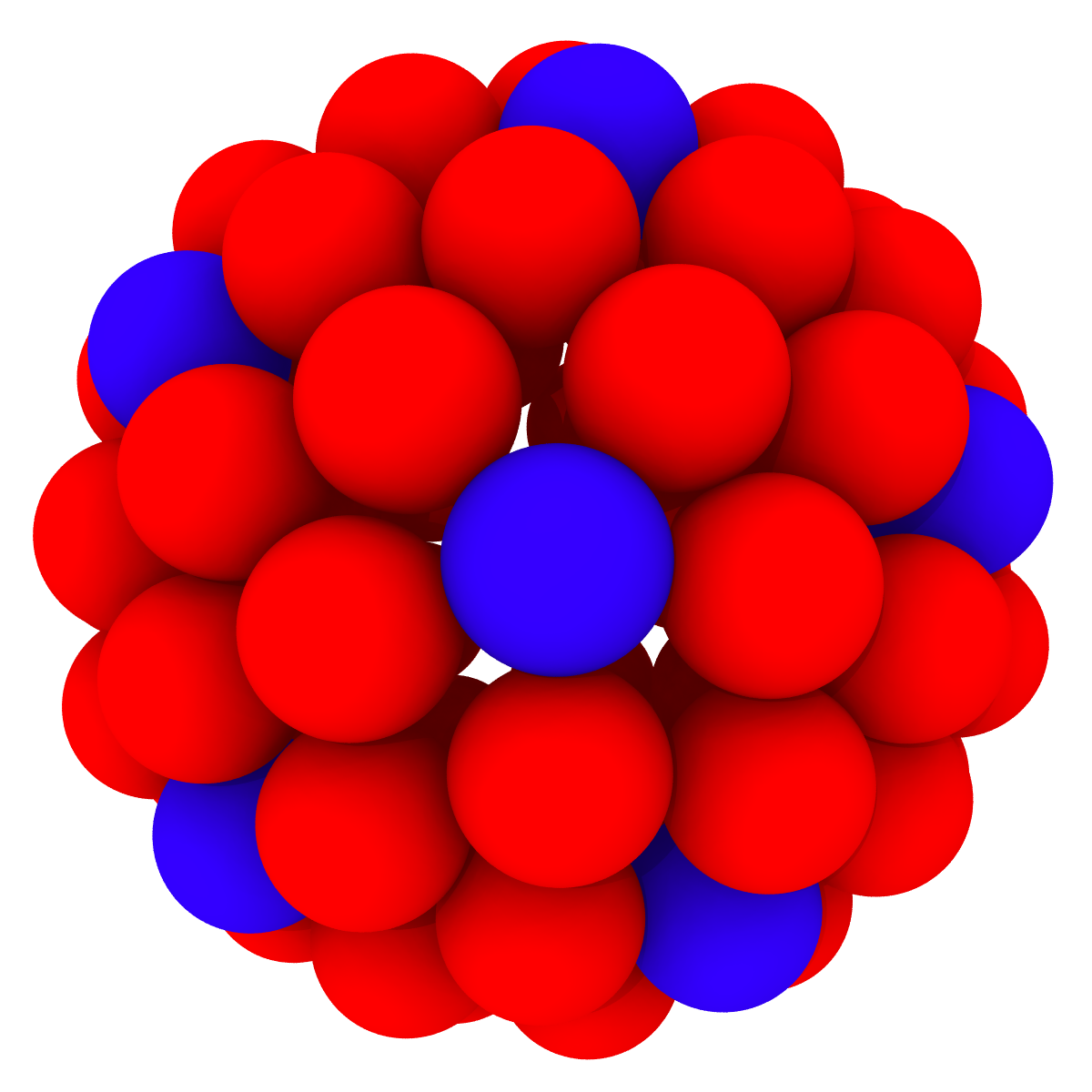}
    \caption{\label{fig:N72-min-ico}}
  \end{subfigure}
  \begin{subfigure}[t]{0.18\textwidth}
    \includegraphics[width=\textwidth]{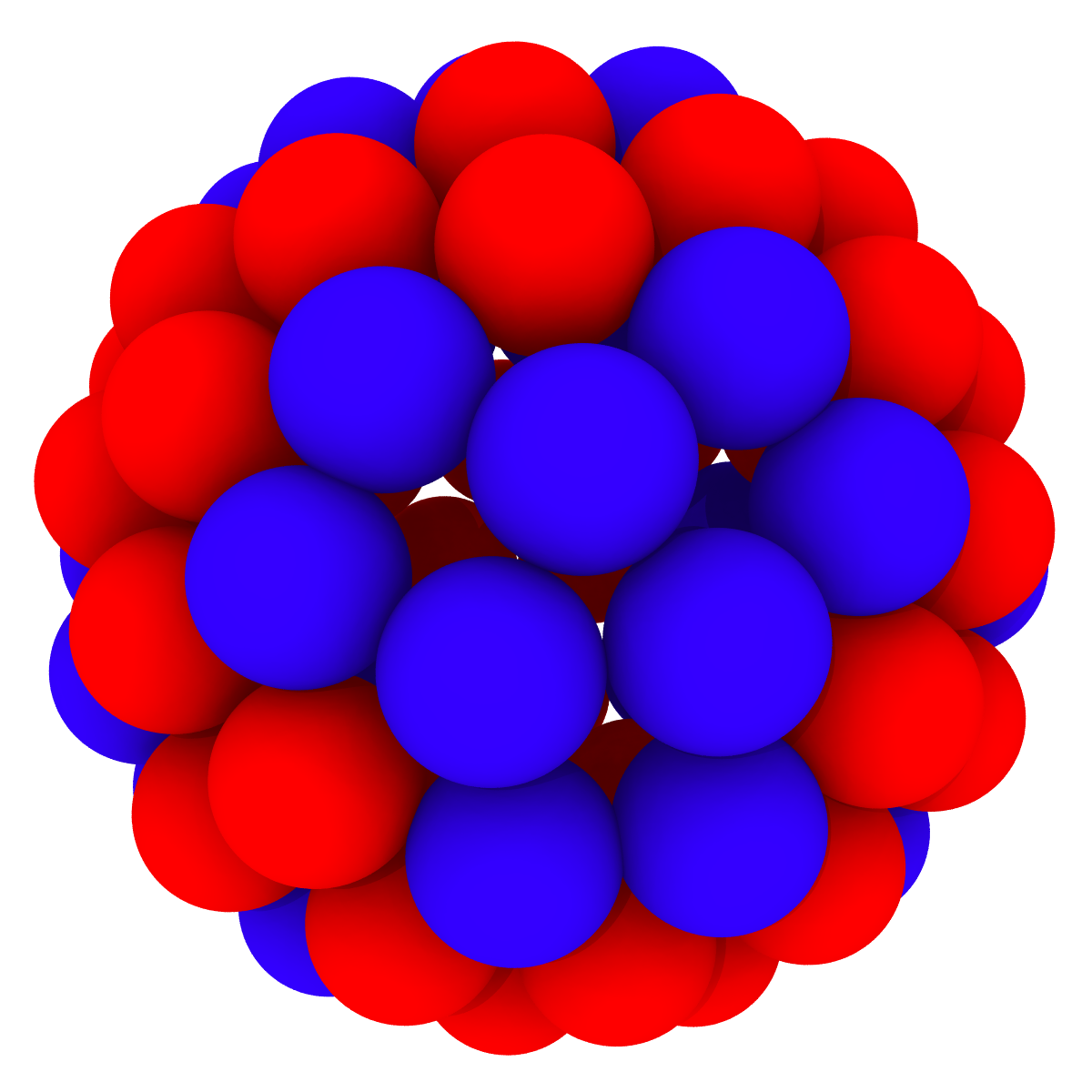}
    \caption{\label{fig:N72-min-weird}}
  \end{subfigure}
  \begin{subfigure}[t]{0.18\textwidth}
    \includegraphics[width=\textwidth]{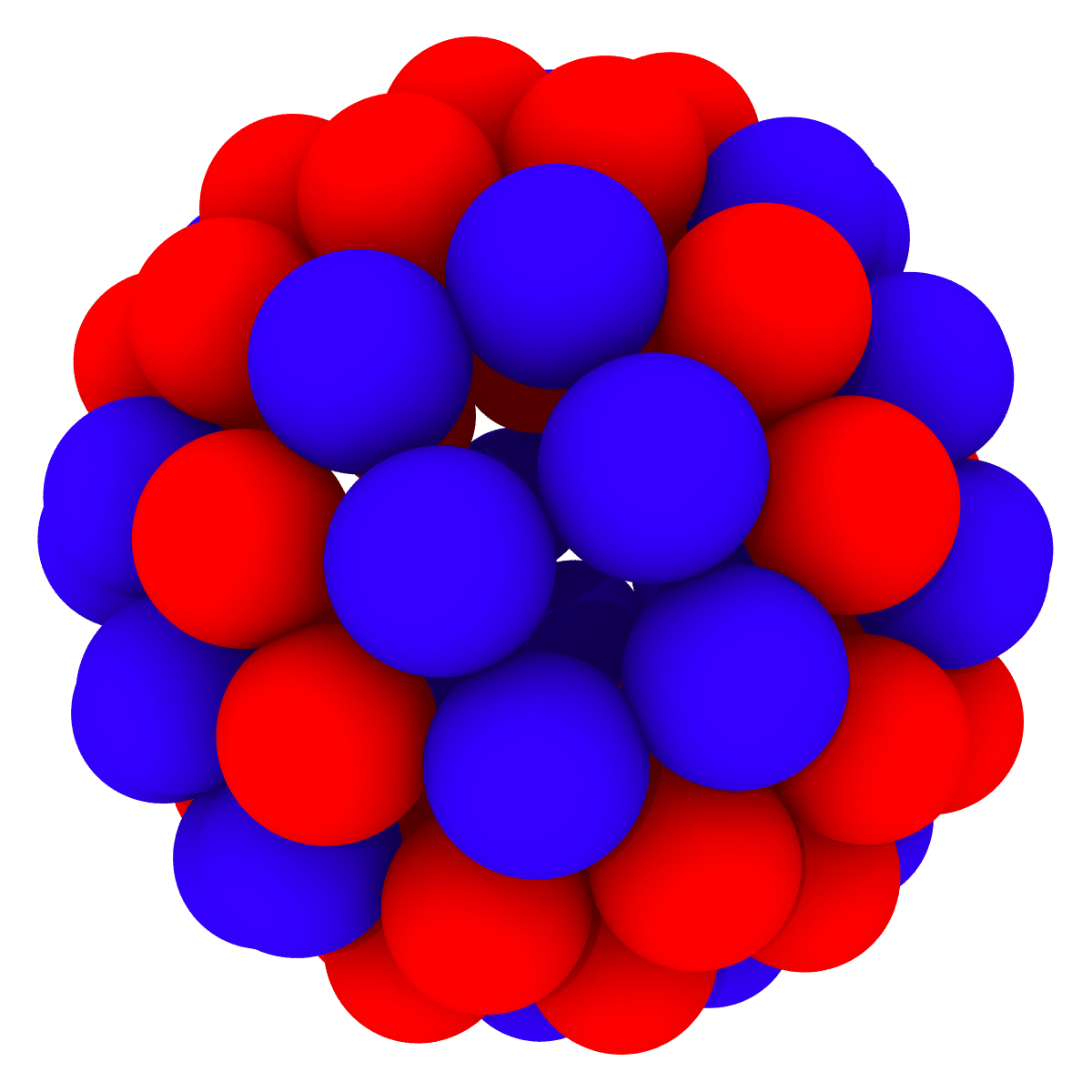}
    \caption{\label{fig:N72-min-other}}
  \end{subfigure} \\
  \caption{\colornote{}The five Lennard-Jones packings for $N=72$ with the lowest potential energy found using the GMIN program \cite{gmin} (a): $D_{5h}$ packing with energy per particle $U/N = -3.0564\epsilon,$ (b): $D_3$ packing with $U/N = -3.0559\epsilon,$ (c): icosahedral packing with $U/N = -3.0548\epsilon,$ (d): tetrahedral packing with $U/N = -3.04636\epsilon$ and (e): packing with two times three rectangular patches that wrap around the sphere similar to the seam on a baseball, with $U/N = -3.04630\epsilon.$ Top and bottom show different orientations. The colour coding indicates the coordination numbers five (blue) or six (red). \label{fig:struct-mismatches-72}}
\end{figure}

\begin{figure}[htb]
  \centering
  \includegraphics[width=0.8\textwidth]{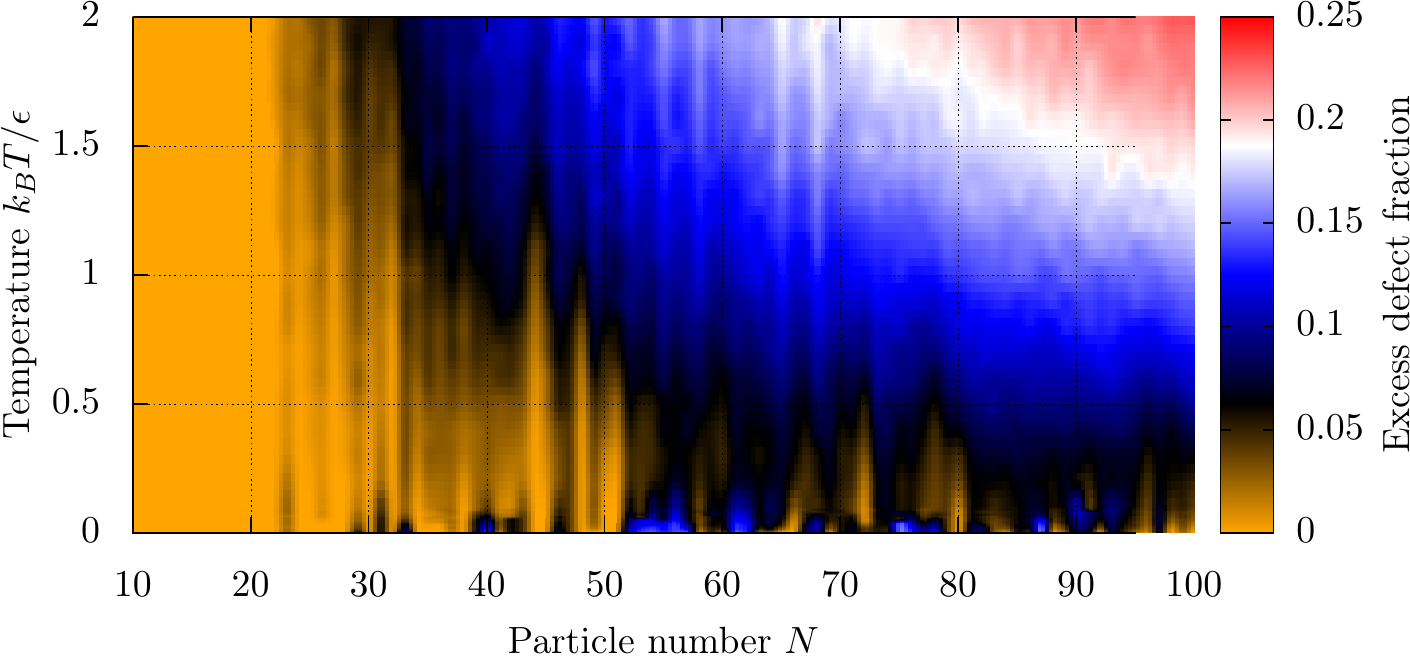}
  \caption{\colornote{}Excess defect fraction for $N=10$ to $N=100$ Lennard-Jones particles determined by means of the convex hull, as described in Section \ref{sec:methods}. \label{fig:lj-hull-defects}}
\end{figure}

\begin{figure}[htb]
  \centering
  \includegraphics[width=0.8\textwidth]{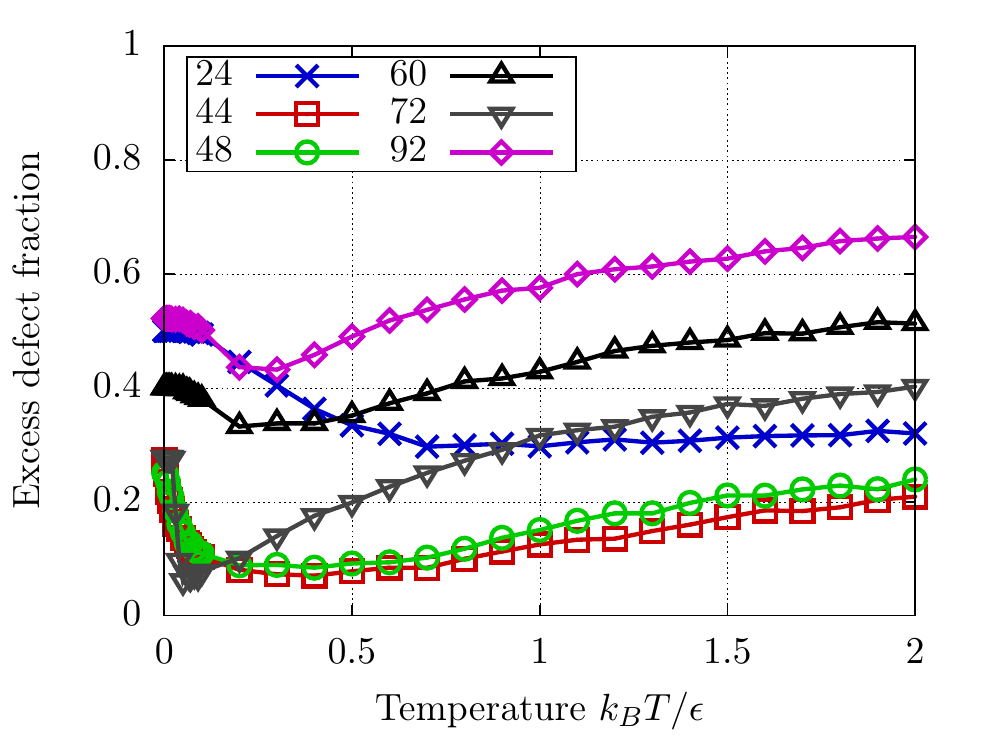}
  \caption{\colornote{}Excess defect fraction for $N=38,~44,~48$, 60, 72 and 92. Lennard-Jones particles as function of temperature. Note the clear re-entrance of excess defects for with increasing temperature for $N\neq 24.$ \label{fig:lj-some-defects}}
\end{figure}

Typically, the number of excess defects increases with temperature. Remarkably, however, for certain particle numbers we observe additional defects in the ground state and a non-monotonic dependence of the number of defects as a function of temperature, most notably for $N=44,$ 48 and 72. $N=72,$ is particularly interesting because one might expect the minimum energy structure to be a $T=7$ icosahedron. While the icosahedron is a low energy minimum, it turns out there are two more packings with a lower potential energy, namely, a $D_{5h}$ structure and a $D_3$ structure, as well as two additional packings with a slightly higher potential energy, one of which exhibits tetrahedral symmetry. We present all of them in Fig. \ref{fig:struct-mismatches-72}.
Apart from the icosahedral structure \ref{fig:N72-min-ico}, they all exhibit clusters of point defects. The two lowest minima have square arrangements of particles. From this result we can conclude that for $N=72$ an icosahedral packing is stabilised entropically rather than energetically. We shall demonstrate that this is indeed the case in Section \ref{sec:lj-free-energy}. Note that there are other particle numbers for which excess defects disappear at intermediate temperatures, \eg{} $N=24$, 44, 48, 60 and 90. For all these particle numbers but $N=24,$ excess defects reappear at higher temperatures. The excess defect fraction for these particle numbers is plotted as function of temperature in figure \ref{fig:lj-some-defects}.

Finally, we consider the same phase diagram but now with the coordination number determined by the convex hull, shown in Fig. \ref{fig:lj-hull-defects}. Again we find that for certain particle numbers, excess defects appear at zero temperature, disappear at intermediate temperatures, then reappear at higher temperatures. 
This observation suggests that our findings are indeed robust.
However, details of the excess defect landscapes calculated from the two methods do vary quite significantly.

For the distance criterion the number of excess defects for a given temperature does not seem to follow a clear trend as a function of the number of particles. However, for the Voronoi tesselations we see a gradual increase in the number of excess defects with increasing $N$ at fixed temperatures $T > 1~\epsilon/k_B.$ Furthermore, the total number of excess defects in this construction is significantly smaller over the entire temperature range. 
For  the Voronoi tesselation the largest excess defect fraction is only $0.25,$ whereas that for the distance criterion it is about $0.8.$ This difference is explained by the fact that at high temperatures, the particles are effectively a liquid and there are large fluctuations in inter-particle distances. These large fluctuations lead to a considerable fraction of particles that have other than six nearest neighbours. The convex hull is not sensitive at all to the inter-particle distance, and thus does not reach these large values.

\begin{figure}[htb]
  \centering
  \begin{subfigure}[t]{0.23\textwidth}
    \includegraphics[width=\textwidth]{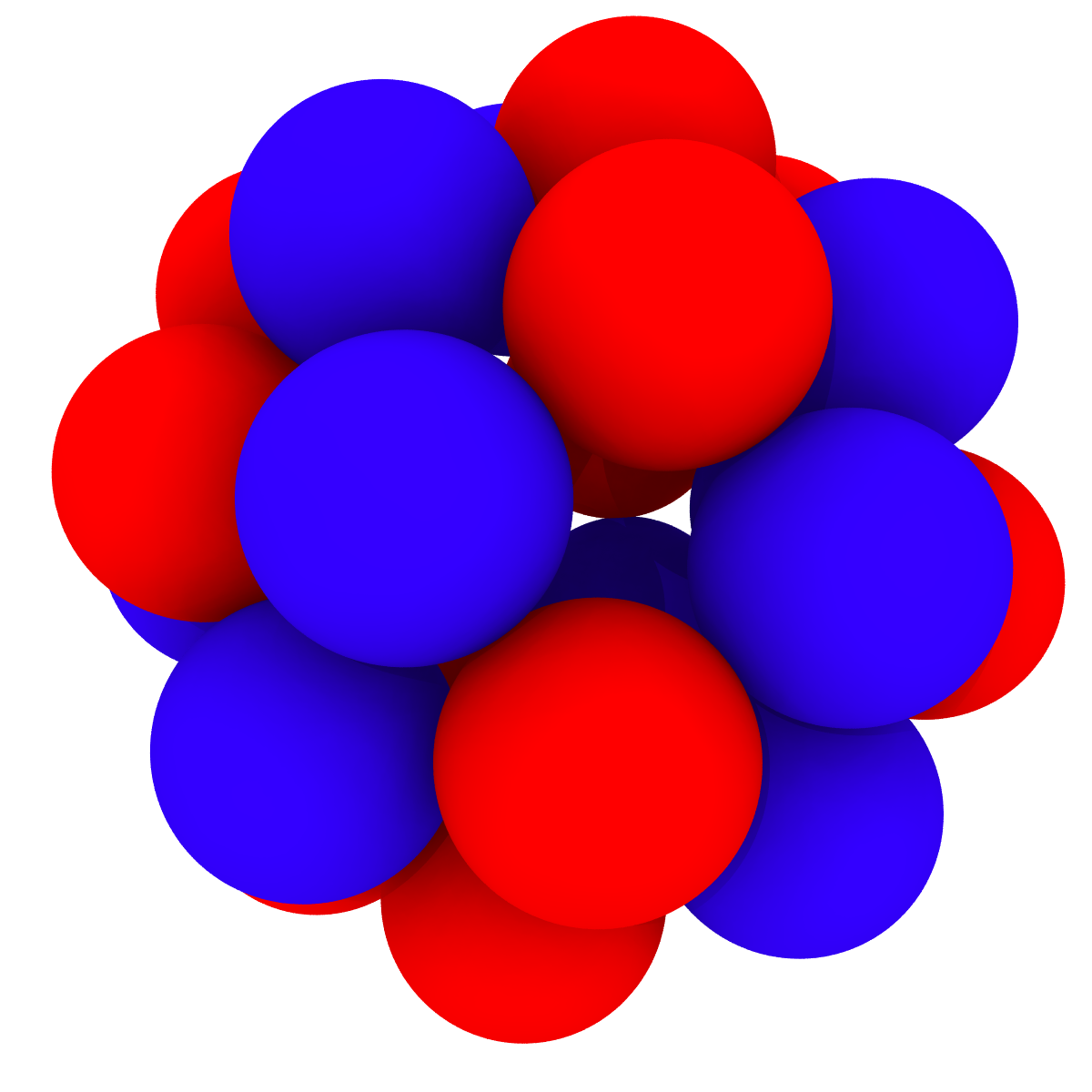}
    \caption*{(a) convex hull}
  \end{subfigure}
  \begin{subfigure}[t]{0.23\textwidth}
    \includegraphics[width=\textwidth]{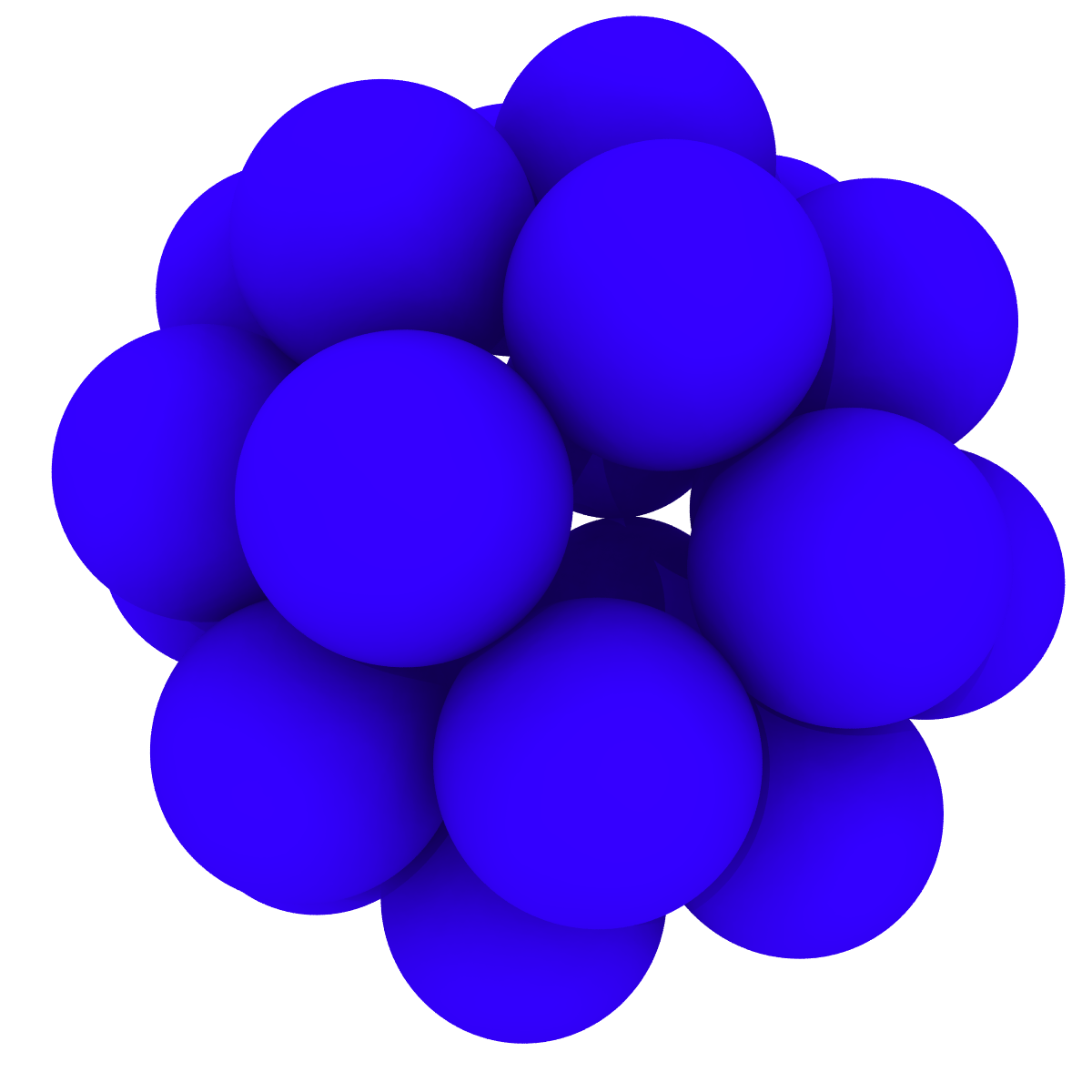}
    \caption*{(a) distance}
  \end{subfigure}
  \begin{subfigure}[t]{0.23\textwidth}
    \includegraphics[width=\textwidth]{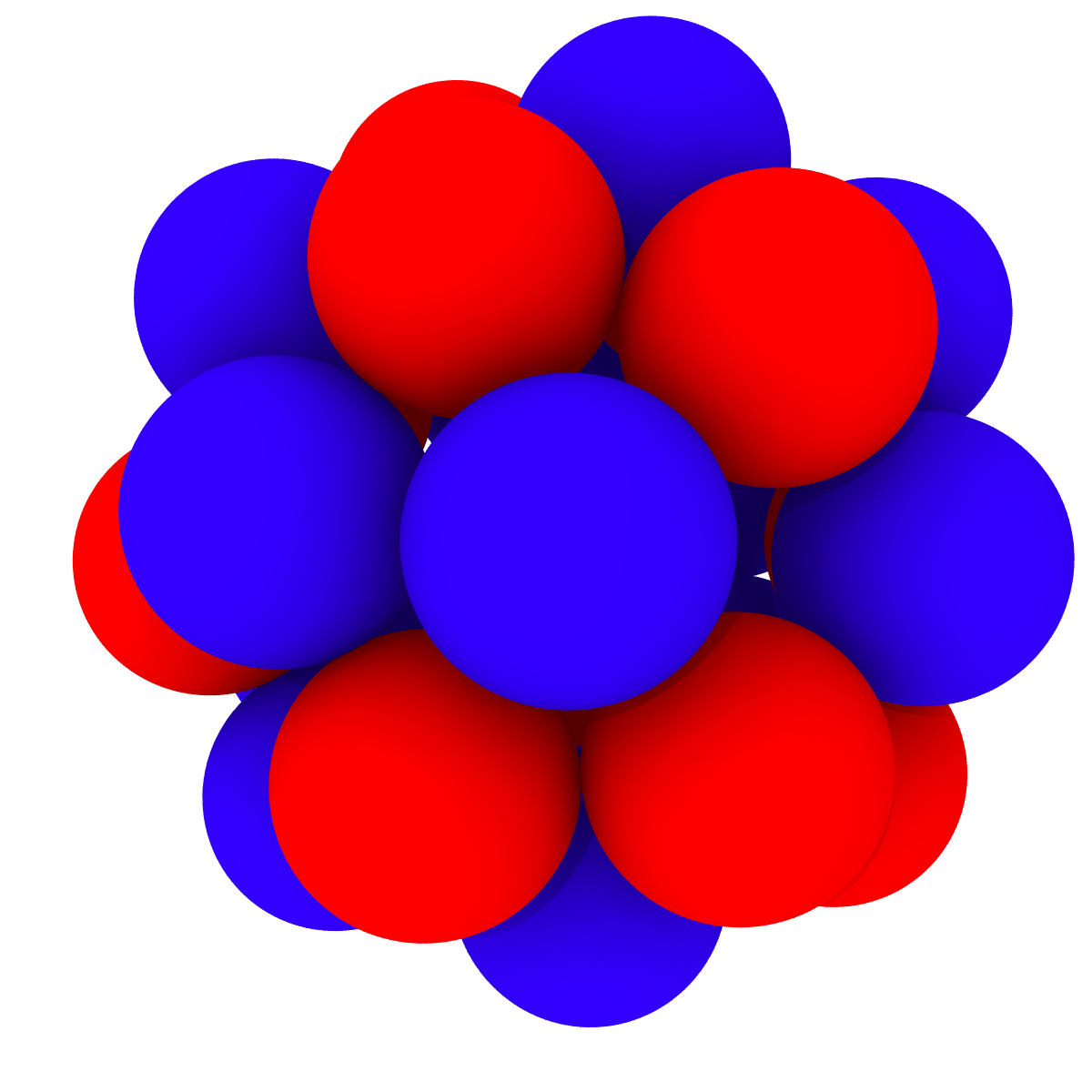}
    \caption*{(b) convex hull}
  \end{subfigure}
  \begin{subfigure}[t]{0.23\textwidth}
    \includegraphics[width=\textwidth]{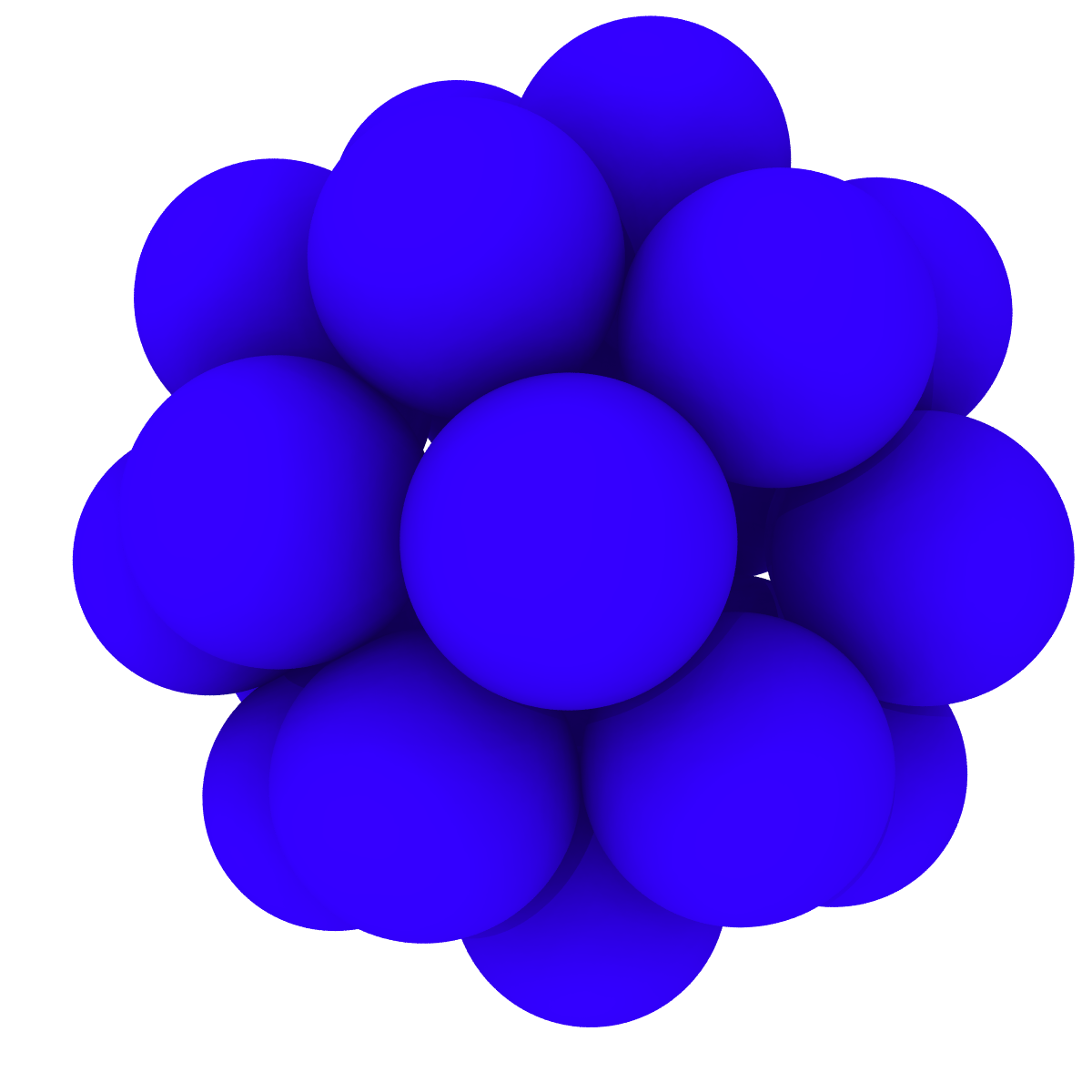}
    \caption*{(b) distance}
  \end{subfigure}
  \caption{\colornote{}Global potential energy minimum for $N=24$ Lennard-Jones particles, shown from two vantage points (a) and (b), obtained using the GMIN program.\cite{gmin} The colour indicates the number of nearest neighbours. Red  particles have 6 and blue 5 nearest neighbours, as identified by the convex hull and the distance criterion.
    \label{fig:struct-mismatches-24}
  }
\end{figure}

Using the Voronoi tesselation, many packings do not have any excess defects for a wide temperature range. This is because the convex hull identifies many particles as having six neighbours, even when the separation distance between them is large. To assess which method gives more insight, we specifically consider the octahedral packing \cite{zandi-2004} for $N=24$ in Fig. \ref{fig:struct-mismatches-24}, where we colour the coordination of the particles according to the Voronoi tesselation and the distance criterion. Each particle in the packing plays an equivalent role, as they are all at the corner of a square arrangement and touch five other particles. Nevertheless, the convex hull arbitrarily assigns six nearest neighbours to some of the particles. In view of this result, we feel the distance criterion to be a better way to determine the coordination number of the particles, and we use it in the remainder of this paper.

\section{Defects near the ground state}
\label{sec:lj-free-energy}

As we have seen in Section \ref{sec:defects-lj}, some particle numbers produce packings that exhibit excess defects at very low temperatures. For two thirds of the particle numbers considered, the number of excess defects obtained for $T=0$ by means of basin-hopping is equal to the number of excess defects at the lowest non-zero temperature result from our Langevin dynamics simulation ($T = 0.001~\epsilon/k_B$). This correspondence suggests that  these packings are not the result of kinetic trapping but are energetically stabilised. Some particle numbers, however, exhibit a discrepancy between the two approaches.

The even $N$ for which there was a minor discrepancy in the excess defect fraction between these two simulations were $N=28, 30, 50, 58, 74, 94,$ and $98.$ For these packings, the particles fluctuate between different low energy structures even at this low temperature, and therefore the average number of excess defects does not exactly match the number of excess defects in the global minimum. The largest relative deviation in the excess defect fraction between the two is $0.14\%$ for $N=30.$ From this result we conclude that if we would go to even lower temperatures, we would get the right structures because the global minimum dominates. We have not pursued this limit further on account of the very long equilibration times required for proper sampling.

For the odd particle numbers, we see similar discrepancies, namely for particle numbers $N=37,$ 39, 41, 43, 47, 51, 55, 59, 73, 79, 85 and all odd $N \geq 89,$ where the largest discrepancy in the excess defect fraction amounts to $0.12\%.$ Again, at the lowest non-zero temperature tested, the particle packing fluctuates between different symmetries, where the dominant structure is the global minimum. For all other odd and even $N,$ we found no discrepancies between the two methods.


For some particle numbers that exhibit excess defects in the low temperature regime, we find that these defects disappear at intermediate temperatures and reappear at higher temperatures. This effect occurs for even $N=28,$ 40, 42, 46, 60, 62, 64, 68, 72, 76 and 86, and for odd $N=37,$ 39, 41, 61, 71, 91 and $97.$
To investigate this unexpected behaviour we focus attention on $N=72$ particles, for which we know that the lowest temperature Langevin dynamics packing coincides with the zero temperature basin-hopping result. Apart from the global minimum, basin-hopping finds four additional local potential energy minima with a significantly lower potential energy than the other local minima ($1.4\%$ difference). The differences in potential energy between the five lowest energy packings are very small ($<0.04\%$). Recall that these minimum energy structures are shown in Fig. \ref{fig:struct-mismatches-72}.

In order of increasing potential energy, the symmetries of these packings are icosahedral, tetrahedral, and finally a packing consisting of two domains containing three rectangular patches that wrap around each other, similar to a baseball pattern. Of these five packings, only those that correspond to the lowest three potential energy minima, (a), (b) and (c) in Fig. \ref{fig:struct-mismatches-72}, are observed in our LD simulations at low but non-zero temperatures, indicating that either the kinetic barrier between these three states and the other two is too large, or that the free energy difference destabilises the two packings with higher potential energy. Taking into consideration the contribution of the potential energy to the Boltzmann weight of a configuration, in particular near zero temperature, this last explanation seems plausible. For these low potential energy packings we present the ratios of the  calculated Boltzmann factors for six temperatures in Table \ref{tab:boltzmann-factors}. We calculated these Boltzmann factors from the potential energies of the packings obtained by means of basin-hopping, given in the caption to Fig. \ref{fig:struct-mismatches-72}. From Table \ref{tab:boltzmann-factors} becomes clear that at very low temperatures the potential energy differences are amplified and that this is what destabilises the tetrahedral and baseball packings.

\begin{table}[b]
  \centering
  \caption{Estimated relative probabilies of observing a $D_3$ (Fig. \ref{fig:N72-min-D3}), icosahedral (ico, Fig. \ref{fig:N72-min-ico}), tetrahedral (tetra, Fig. \ref{fig:N72-min-weird}) or a packing with two domains with three rectangular patches (rect, Fig. \ref{fig:N72-min-other}) compared to that of finding $D_{5h}$ (Fig. \ref{fig:N72-min-D5h}), using the Boltzmann weight of the respective calculated potential energy. \label{tab:boltzmann-factors}}
  \begin{tabular}{c|cccccc}
    $k_B T / \epsilon$    & 0.001 & 0.01 & 0.02 & 0.03 & 0.04 & 0.05 \\  \hline
    $P(D_{3}) / P(D_{5h})$ & 0.556 & 0.943  & 0.971 & 0.980 &  0.985 & 0.988 \\
    $P(\text{ico}) / P(D_{5h})$ & 0.199 & 0.851& 0.922&0.948 &0.960 &0.968  \\
    $P(\text{tetra}) / P(D_{5h})$ & $4.18~10^{-5}$  &0.365 & 0.604&0.715 &0.777 &0.817  \\
    $P(\text{rect}) / P(D_{5h})$ &  $3.95~10^{-5}$  &0.363 & 0.602&0.713 &0.776 &0.816  
  \end{tabular}
\end{table}

In order to quantify the free energy rather than the potential energy differences between the three packings found in our dynamics simulations, we determine the frequency of occurrence of the different packings, as outlined in Section \ref{sec:methods}. To verify ergodicity, we keep track as a function of time the normalised frequencies of each packing, and ascertain that they reach a steady state value. Additionally we verify that the frequencies obtained from different initial packings have converged to each other. In particular, we prepare our systems in the $D_{5h},~D_3$ and icosahedral symmetries, corresponding to the three lowest-energy packings in Fig. \ref{fig:struct-mismatches-72}. For $T > 0.03~ \epsilon/k_B$ ergodicity seems to hold. The details of the identification and sampling of the packings are discussed in SI 3.

\begin{figure}
  \centering
  \includegraphics[width=0.8\textwidth]{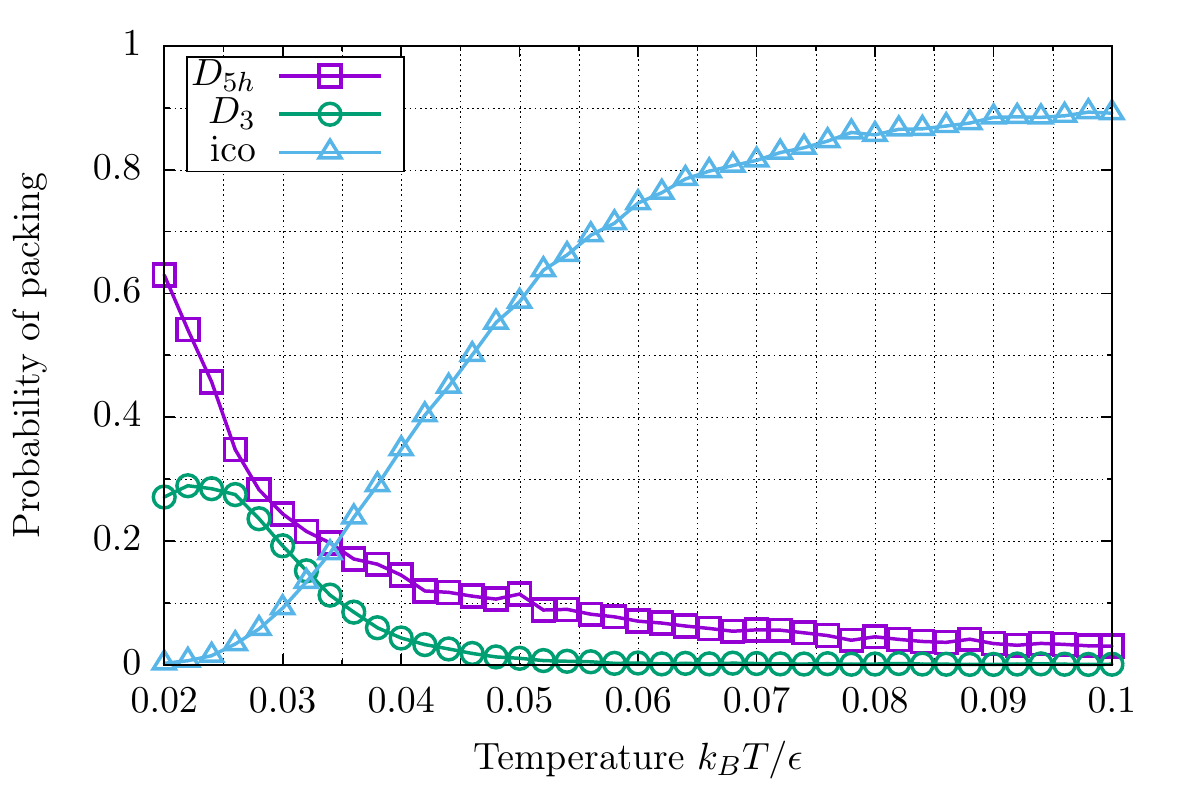}
  \caption{\colornote{}Probability of encountering an icosahedral, $D_{5h}$ or $D_3$ packing with $N=72$ Lennard-Jones particles on a sphere with radius $R=2.55037 r_0,$ where $r_0$ is the equilibrium spacing of the pair potential, as function of the dimensionless temperature $k_B T / \epsilon.$ The frequencies do not sum to unity because for some time frames the packing could not be identified. \label{fig:freqs}}
\end{figure}

In Fig. \ref{fig:freqs} we show the frequencies at which the different packings occur as a function of temperature. Note that at low temperatures, the low potential energy packings $D_3$ and $D_{5h}$ are energetically stabilised, while the icosahedral packing is completely suppressed. At higher temperatures, the icosahedral packing becomes more and more dominant, while the $D_{5h}$ and $D_3$ packings become entropically suppressed. Basin-hopping predicts a $D_{5h}$ packing for the global potential energy minimum, which is consistent with the trend shown in Fig. \ref{fig:freqs}, but reliable data for the temperatures in between $T=0$ and $0.03~\epsilon/k_B$ are difficult to obtain due to the increased simulation time needed for proper sampling. Thus, while for $T\leq 0.03~\epsilon/k_B$ the trend seems to be consistent with the basin-hopping calculations, the exact values for the frequencies might not be that reliable. For $T > 0.03~\epsilon/k_B$ a clear steady-state was reached that converged for all three initial packings, and thus we presume these data to be reliable.

\begin{figure}
  \centering
  \includegraphics[width=0.8\textwidth]{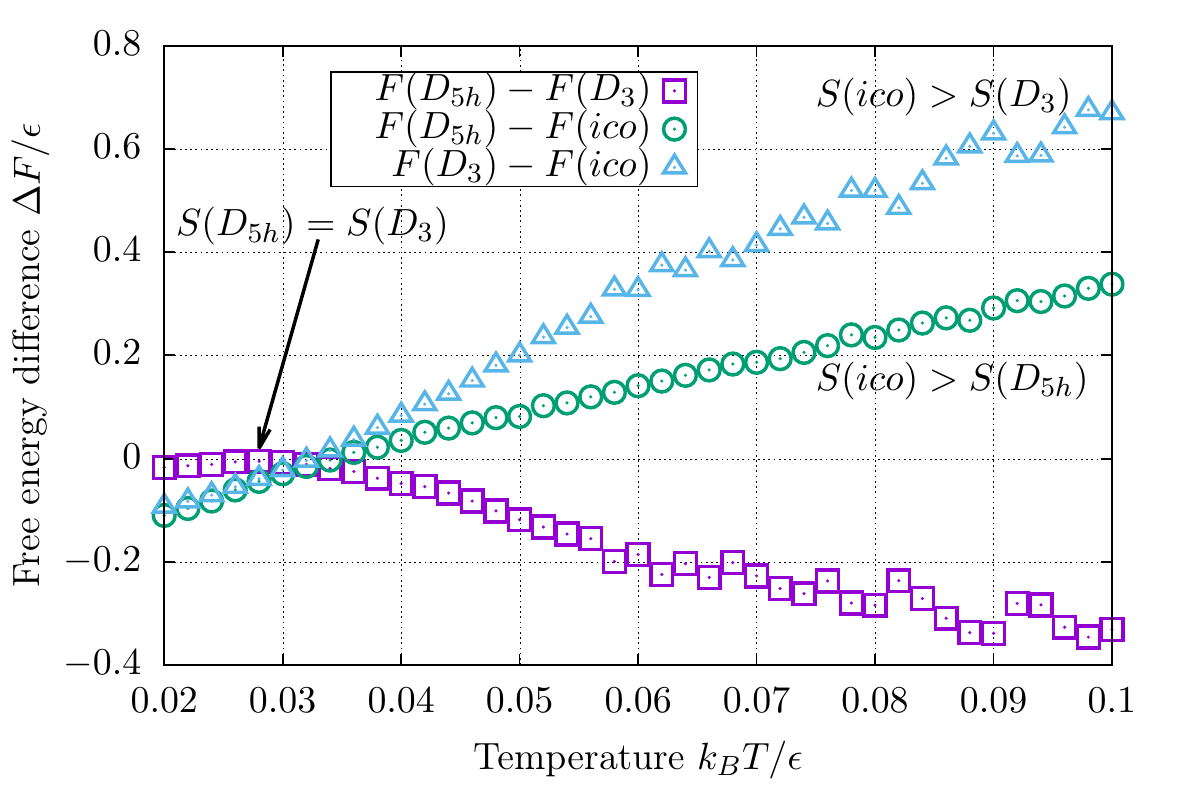}
  \caption{\colornote{}Free energy differences between packings of $N=72$ particles on a radius $R=2.55037 r_0.$ At low temperatures the $D_3$ and $D_{5h}$ packings are nearly equal in free energy, but an increasing importance of entropy destabilises the $D_3$ packing more than the $D_{5h}$ at higher $T.$ Both the $D_{5h}$ and $D_3$ packing are destabilised at higher $k_B T$ in favour of the icosahedral packing. At $k_B T \approx 0.032\epsilon$ the three packings appear to be equally probable.    \label{fig:dFs}}
\end{figure}

Using the relative occurrence frequencies of the different symmetries we extract free energy differences, presuming ergodicity, from the associated Boltzmann weights. In Fig. \ref{fig:dFs} we plot these free energy differences, from which we immediately see that at around $T \approx 0.032~ \epsilon/k_B$ all three packings are equally likely, and that above that temperature the free energy of an icosahedral packing is the lowest. Thus, above $T = 0.032~ \epsilon/k_B,$ we expect to see predominantly the icosahedral packing, which is consistent with Fig. \ref{fig:freqs}.

Furthermore, we can determine the entropy differences by calculating the slope of $\Delta F,$ since $S = -(\partial F/\partial T)_{N,R},$ evaluated at constant particle number $N$ and spherical template radius $R.$ This analysis immediately reveals that the entropy of an icosahedral packing is larger than that of both the $D_{5h}$ and $D_3$ packings, as the slopes of $F(D_{5h}) - F(ico)$ and $F(D_3)-F(ico)$ are positive for the entire temperature range probed. Also note that the entropy of the $D_{5h}$ packing is larger than that of the $D_3$ packing for most temperatures, as $F(D_{5h}) - F(D_3)$ has a negative slope for $T>0.025~\epsilon/k_B.$ Hence, at higher temperatures, the icosahedral packing is favoured over both the $D_{5h}$ and $D_3$ packings due to its higher entropy, while at low temperatures the $D_{5h}$ packing is preferred due to its low potential energy and the fact that its entropy is higher than that of the $D_3$ packing.

For even larger temperatures $T > 0.1~\epsilon/k_B$ the icosahedral packing becomes less stable because of the emergence of thermally excited excess defects, as is clear from Fig. \ref{fig:lj-dist-defects}. For this range of temperatures we did not explicitly obtain a free energy difference because we find many different packings, none of which seem to be clear potential energy minima.
Thes results confirm, not surprisingly, that the equilibrium packings of particles on a curved surface are not just a result of potential energy minimisation but rather of free energy minimisation. Finally, it is clear that on curved surfaces, additional point defects can actually lower the potential energy, and are thus energetically stabilised. Although we have only explicitly shown this for $N=72$ particles, we hypothesise that the same effect occurs for other particle numbers that exhibit additional defects in the ground state, which disappear for intermediate temperatures, \eg{}, $N=60$ and $N=92.$

Now that we have shown that the temperature, or, equivalently, the interaction strength, plays a crucial role in stabilising different packings, we turn to the role of the range of attraction of the interaction potential.

\section{Morse defect landscape}
\label{sec:defects-morse}

In the previous section we saw that for Lennard-Jones particles there exist energetically stabilised defects at low temperatures. Furthermore, we found that icosahedral packings are stabilised energetically for $N=32$ but only entropically for $N=72.$ Since a shorter ranged potential is a more realistic model in the context of colloidosomes and virus capsomeres, it is of interest to see how robust our findings are if we reduce the effective range of the interaction potential. We set the range parameter $\alpha=60/r_0$, as discussed in Section \ref{sec:methods}, and again determined the excess point defect landscape as a function of particle number $N$ and temperature $T.$

\begin{figure}[htb]
  \centering
  \includegraphics[width=0.8\textwidth]{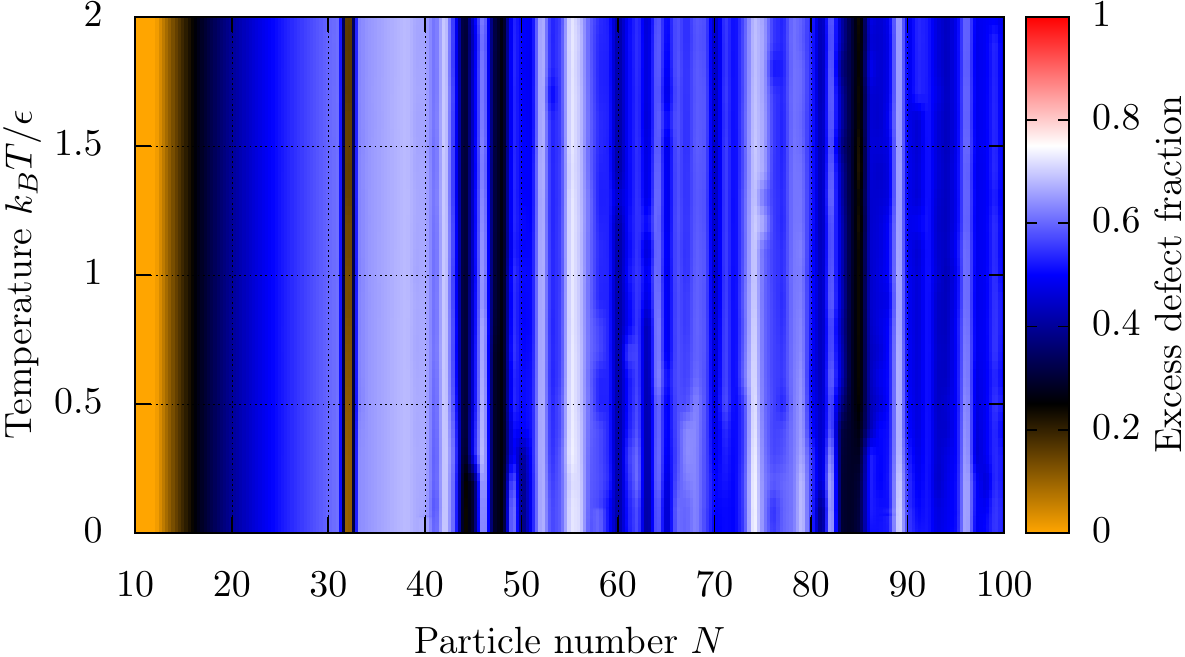}
  \caption{\colornote{}Temperature dependence of the excess defect fraction using the distance criterion for $N=10$ to $N=100$ Morse particles with a range parameter of $\alpha = 60/r_0.$  \label{fig:morse-dist-defects}}
\end{figure}

In Fig. \ref{fig:morse-dist-defects} we show the defect landscape using the distance criterion. Note that for the Morse potential, $N=32$ has no additional defects in the ground state, indicating that the icosahedral packing is again energetically stabilised. For $N=72,$ however, there is no longer an intermediate temperature range for which the icosahedral packing is thermally stabilised.

\begin{figure}[htb]
  \centering
  \begin{subfigure}[t]{0.25\textwidth}
    \includegraphics[width=\textwidth]{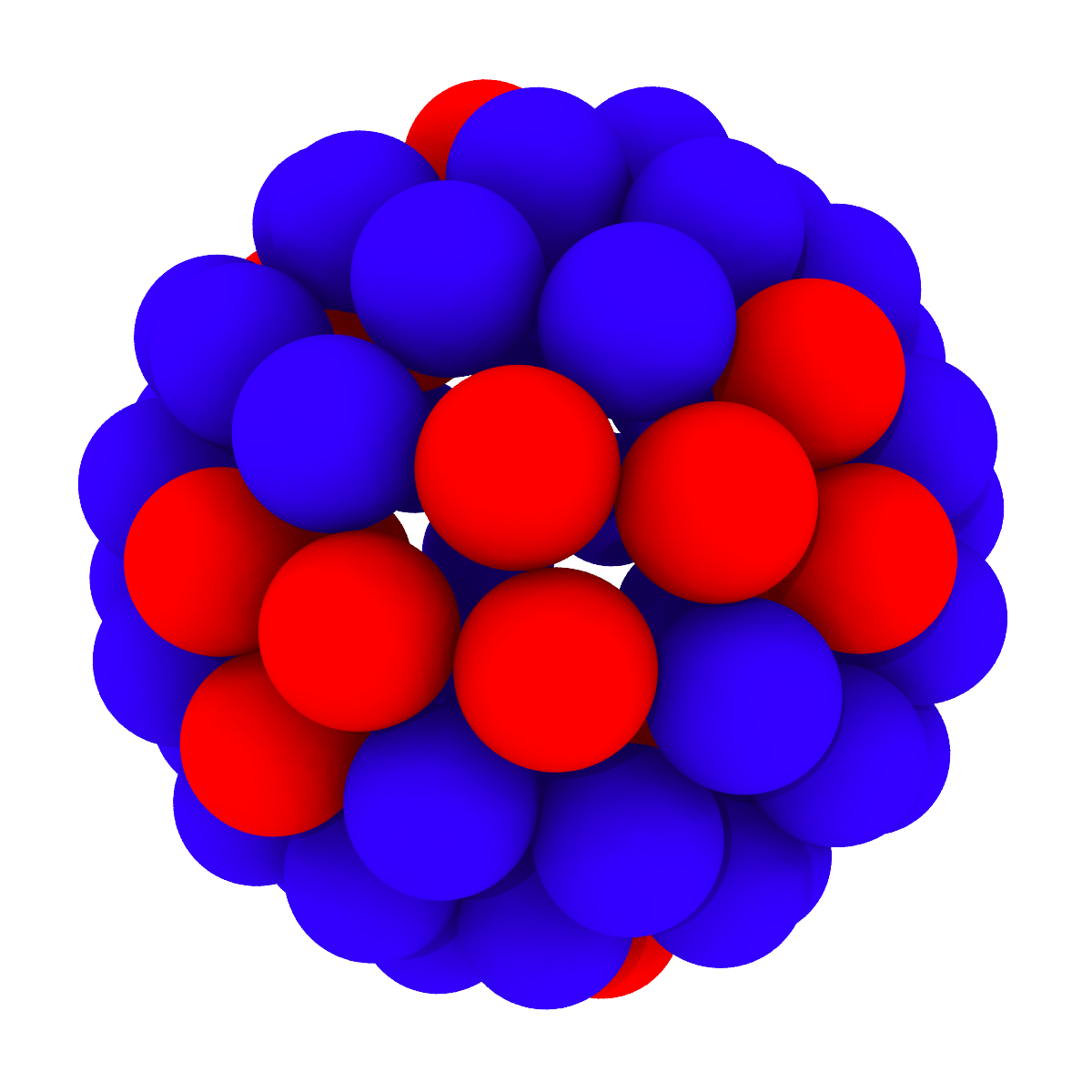}
  \end{subfigure}
  \begin{subfigure}[t]{0.25\textwidth}
    \includegraphics[width=\textwidth]{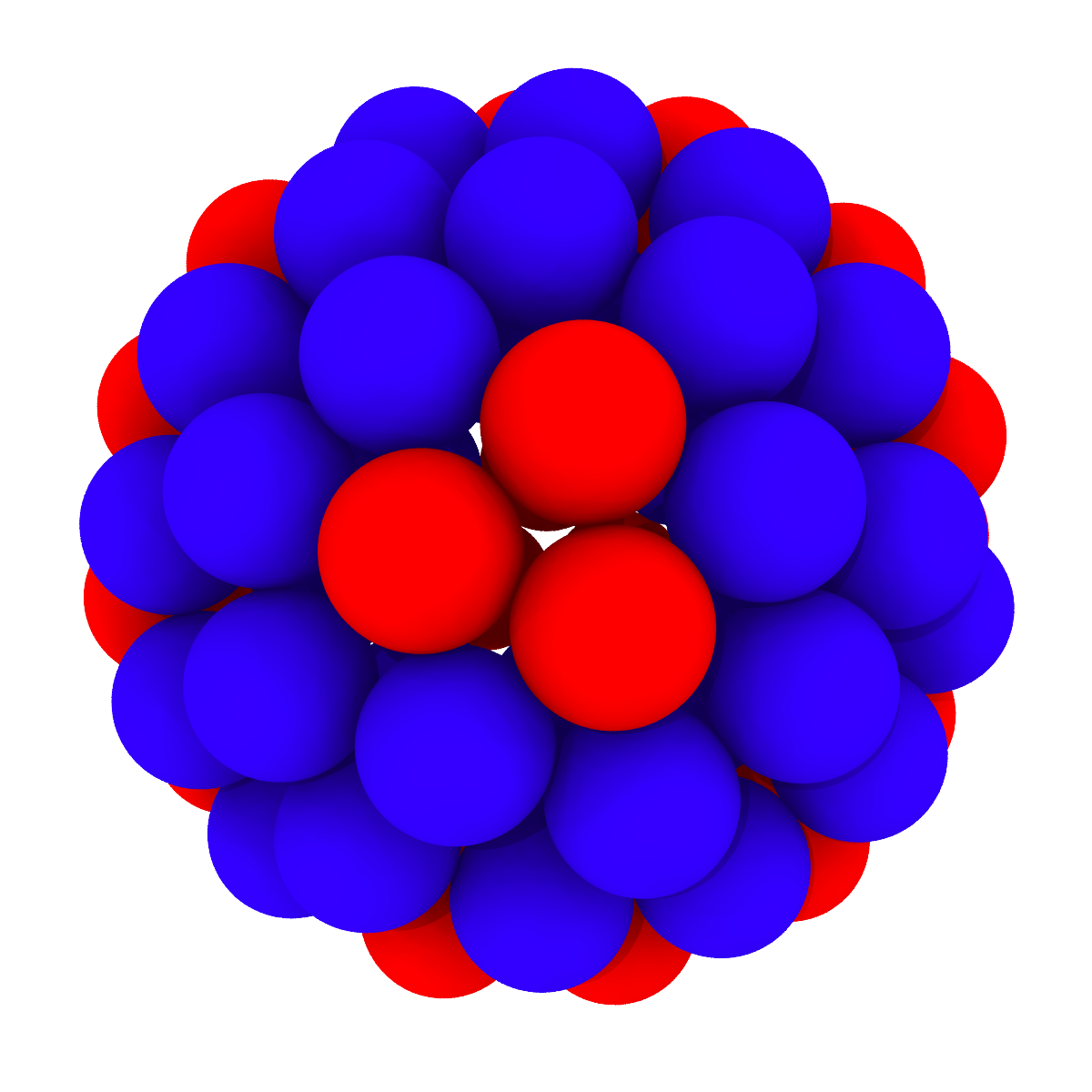}
  \end{subfigure} \\
  \begin{subfigure}[t]{0.25\textwidth}
    \includegraphics[width=\textwidth]{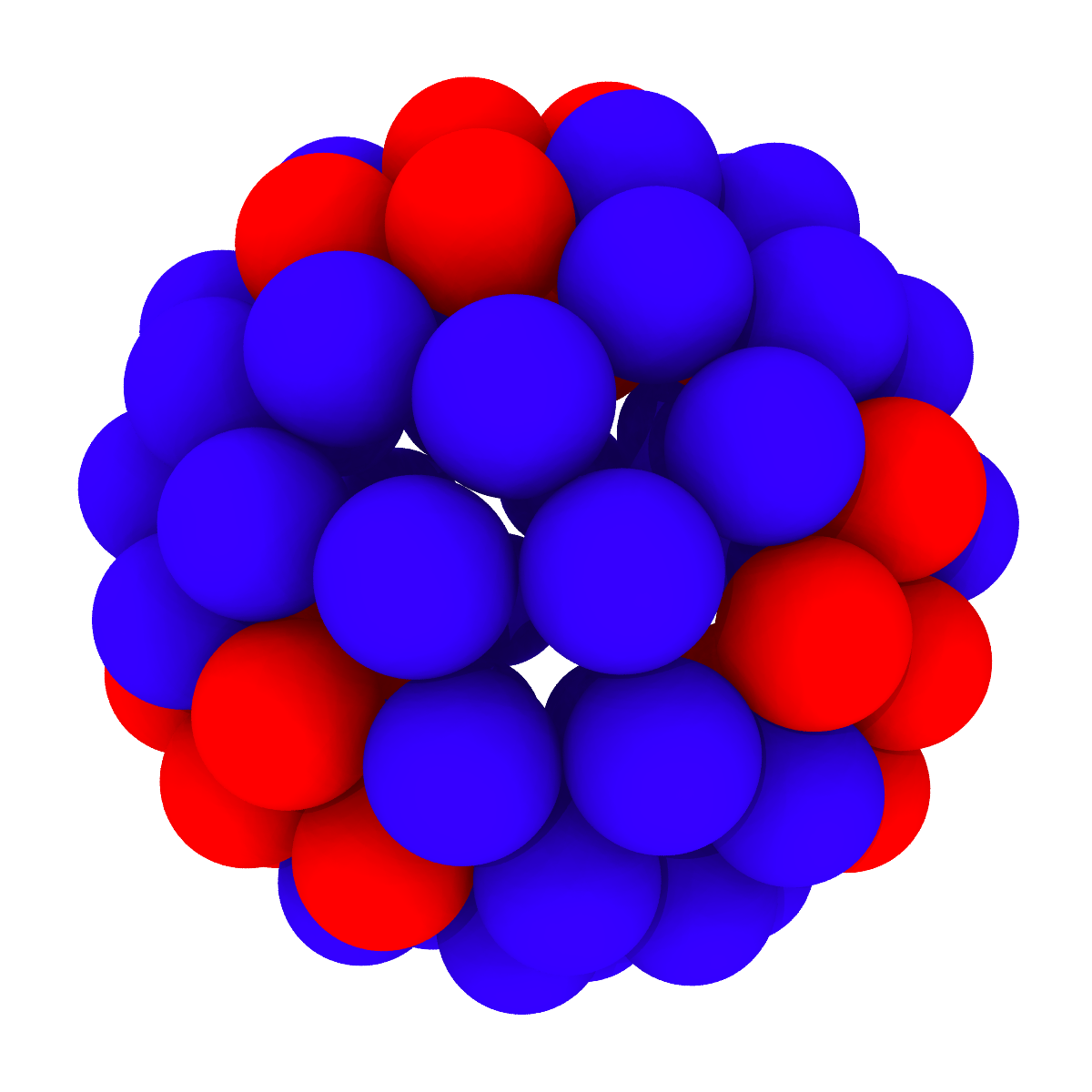}
    \caption{\label{fig:morse-new-minimum-72}}
  \end{subfigure}
  \begin{subfigure}[t]{0.25\textwidth}
    \includegraphics[width=\textwidth]{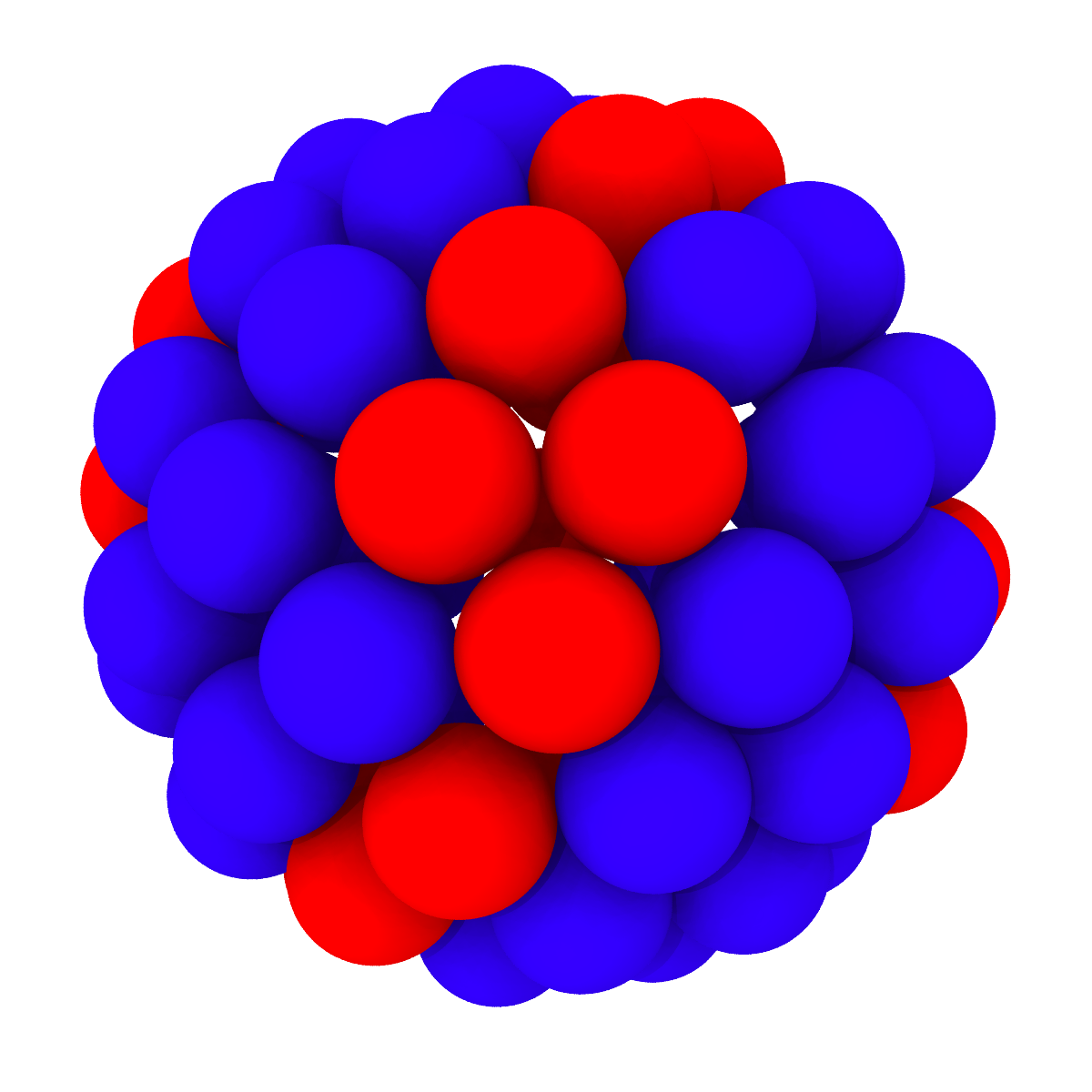}
    \caption{\label{fig:morse-new-minimum2-72}}
  \end{subfigure}
  \caption{The two lowest potential energy packings for $N=72$ particles for a Morse potential with effective range parameter $\alpha = 60 / r_0,$ where $r_0$ is the pair potential equilibrium spacing. Colour codes the  number of nearest neighbours of 5 (blue) or 6 (red). (a): the lowest observed potential energy minimum from two sides (top and bottom), which has an average energy of $U/N = -2.32506 \epsilon.$ (b): The second lowest local potential energy minimum  with $U/N = -2.32348 \epsilon.$ \label{fig:morse-new-minima}}
\end{figure}

The ground state of the $N=72$ packing obtained by basin-hopping consists of three strips of particles with six-fold coordination surrounded by those with five-fold coordination (see Fig. \ref{fig:morse-new-minimum-72}).
The packing corresponding to the second-lowest local potential energy minimum is shown in Fig. \ref{fig:morse-new-minimum2-72}, where the potential energy is $0.068\%$ larger. The other local potential energy minima have significantly higher energies, with the third-lowest having a potential energy 3.4\% larger than the second-lowest. 

In our Langevin dynamics simulations the two packings shown in Fig. \ref{fig:morse-new-minima} are also the most dominant ones. Even at $T = 2~\epsilon/k_B$ the system tends to fluctuate between these two packings, where the second minimum shown in Fig. \ref{fig:morse-new-minimum2-72} only appears very infrequently. Therefore, it seems that for shorter-ranged potentials the energetic penalty is more difficult to overcome by entropy.

From these findings, it seems that a shorter potential range destabilises icosahedral symmetry. For virus capsids this result would imply that, if the capsomeres are all one size, their effective range parameter should be smaller than $\alpha < 60/r_0.$ On the other hand, icosahedral packings can be made more stable by switching between different particle sizes, as discussed by Bruinsma, Zandi \emph{et al.}\cite{bruinsma-2004,zandi-2004} We intend to pursue this question in future work.

Our simulations highlight two major differences between the Lennard-Jones and Morse particle packings.
First, we note that excess defects are barely excited at higher temperatures for particles interacting via the short-ranged Morse potential. This is not entirely surprising because the Morse potential is much steeper than the Lennard-Jones potential, implying that at equivalent thermal energies Morse particles have less opportunity for rearrangements.
Second, and perhaps more strikingly, particle numbers $N>32$ that exhibit a local minimum in the potential energy for both potentials correspond to very different arrangements. These features result in different numbers of excess defects for the two potentials for equal particle number and temperature. This analysis confirms that the range of the potential is very important for determining which particle arrangement is the most favourable.

\begin{figure}[htb]
  \centering
  \includegraphics[width=0.8\textwidth]{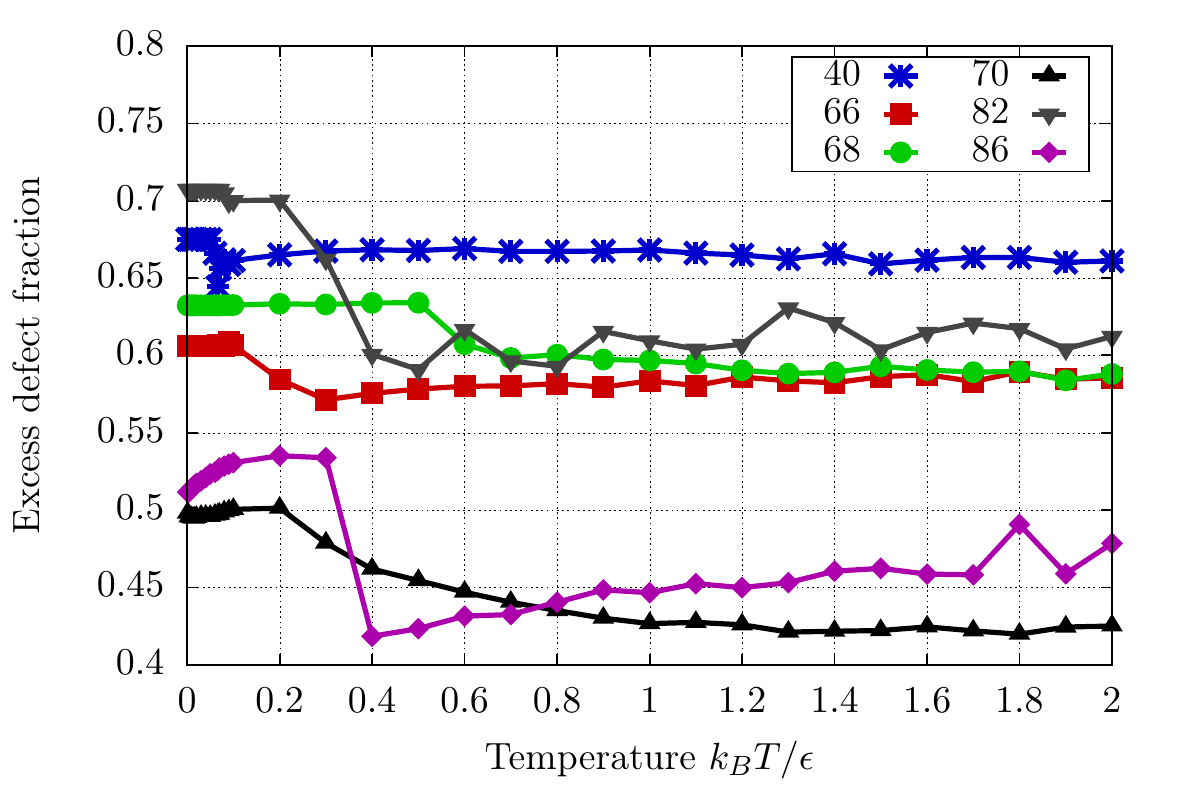}
  \caption{Temperature dependence of the fraction of excess defects for $N=40,$ 66, 68, 70, 82, 86 and 90 Morse particles with range parameter $\alpha = 60/r_0.$  \label{fig:morse-dist-filtered}}
\end{figure}

In spite of this, the similarities to Lennard-Jones particles, Morse particles also exhibit for some $N$ additional defects in the ground state that are energetically stabilised. This is true for $N=40,~66,~68,~70,~82,~86$ and $90$ although for Morse particles this effect is less pronounced, \emph{i.e.}, the range of variation in the excess defect fraction is not as large as for the Lennard-Jones particles. In fact, the range of variation is so small that it is almost indiscernible in Fig. \ref{fig:morse-dist-defects}. Hence, we also plot the excess defect fraction as a function of temperature for the particle numbers quoted in Fig. \ref{fig:morse-dist-filtered}. From the figure, we conclude that these particle numbers show a clear non-monotonic behaviour of the excess defect fraction with increasing temperature, indicating that the defects  at low $T$ for these packings are energetically stabilised, just like the defects we find for the Lennard-Jones packings in Fig. \ref{fig:lj-dist-defects}.

\section{Conclusions}
\label{sec:conclusion}

Inspired by virus capsid and colloidosome assemblies, we have studied by means of computer simulation the packings from $N=10$ to $100$ point particles constrained to a spherical surface. Our aim was to investigate how the optimal particle arrangements are influenced by temperature, or, equivalently, interaction strength, and the range of the interaction potential. These factors have not received extensive attention in the literature, although we find from our simulations that both have a profound impact.
The simulation techniques that we applied involved Langevin dynamics for non-zero temperatures and basin-hopping calculations for determining the global potential energy minima, which confirm that our Langevin simulations are not kinetically trapped at low temperatures.  We have focused mainly on how the number and configuration of point defects, as a measure for the structural stability of packings, vary with temperature. Since at least twelve five-fold point defects are required by geometry, we focus specifically on the number of defects in excess of these twelve.

For $N=12,~32$ and 72 Lennard-Jones particles, we find in the temperature range of $T=0.05 ~\epsilon/kB$ to $T=0.067 ~\epsilon/k_B$ that the equilibrium packing is an icosahedron, consistent with the earlier work of Zandi \emph{et al.}\cite{zandi-2004} Our basin-hopping calculations show that the icosahedral packing is the global potential energy minimum for $N=12$ and 32, but not for 72, for which the global minimum is a $D_{5h}$ packing, in agreement with the results of Voogd.\cite{voogd-thesis} This result is surprising because the $D_{5h}$ packing exhibits additional defects, which apparently have an energetically stabilising effect. Hence, the icosahedral structure for $N=72$ at non-zero temperature must be entropically stabilised. In fact, our simulations suggest that for a fairly large number of particle packings the lowest energy structure exhibits excess defects that, remarkably, disappear when raising the temperature. Of course, at higher temperatures still, defects become thermally excited. For these specific particle numbers the number of defects is a non-monotonic function of temperature, whilst for all others, the number of defects increases with temperature monotonically.

To investigate this kind of ``re-entrant'' behaviour in more detail, we consider $N=72$ Lennard-Jones particles, for which we have explicitly determined free energy differences between the three lowest-energy structures. 
We find that packings with more excess defects have a lower free energy at sufficiently low temperatures, implying that they are energetically favoured over packings with fewer defects. The global potential energy minimum has $D_{5h}$ symmetry. However, our calculations show that the $T=7$ icosahedral packing has a higher entropy than the $D_{5h}$ packing and is therefore thermally stabilised at higher but not too high temperatures.
Therefore the packing of Lennard-Jones particles on curved surfaces is not just governed by minimisation of the potential energy. What is true for $N=72$ seems to be true for many particle numbers, because the symmetries of the associated packings exhibit a strong temperature dependence. On the other hand, the $T=3$ icosahedral symmetry for $N=32$ particles \emph{is} stable over a wide temperature range.

Our main conclusions are not significantly altered if we replace the Lennard-Jones potential by a short-ranged Morse potential, which arguably is more representative of attractive interactions between large molecules or colloidal particles, because it accounts for a larger excluded volume effect.\cite{doye-1995,doye-1996} Again we find that for certain particle numbers, the number of excess defects is a non-monotonic function of the temperature albeit that for most packings we find that it is more difficult to thermally excite additional defects. The latter result implies that packings of particles with a shorter range of attraction are more stable against thermal fluctuations.
Another notable difference between Lennard-Jones and Morse particles is that for equal particle number and temperature, the equilibrium packings may exhibit different symmetries. In particular, this is the case for $N>24,$ with the exception of the $T=3$ icosahedron for $N=32.$

Our calculations suggest that specific predictions for particle geometries on a curved surface depend not only on the strength but also on the exact shape of the interaction potential. Both factors impact upon to what extent temperature is able to affect the competition between particle packings.


\section{Acknowledgements}
S.P. acknowledges the HFSP for funding under grant RGP0017/2012.
This work was partially supported by the National Science Foundation through Grant No. DMR-1310687 (RZ).


\input{paper_new4.bbl}

\clearpage{}


\renewcommand{\thesection}{SI \arabic{section}}
\renewcommand{\thefigure}{SI \arabic{figure}}
\renewcommand{\thetable}{SI \arabic{table}}

\setcounter{section}{0}
\setcounter{figure}{0}
\setcounter{table}{0}

\section*{ Supporting Information for Energetically favoured defects in dense packings of particles on spherical surfaces }

\section{Optimal sphere radius and energy}
In this section we present our results for the optimal sphere radius $R^*$ and the corresponding energies that are described in Section 2. We also show the optimal Lennard-Jones radii and energies reported by Voogd in \cite{voogd-thesis}. They are located in Tab. \ref{tab:energies-radii}. Our values are obtained from simulations in which the radius of the spherical template slowly shrinks over a range estimated from hard disk packings in 2D that presumably contains the optimal radius. Note that our radii for the Lennard-Jones packing do not improve upon Voogd's values, they are in the correct ballpark, with the largest difference in the radius being 2\%. Thus, our strategy provides a good way to obtain a first order estimate for the optimal radius on which more intensive optimisation can be performed.

\begin{center}
  \begin{longtable}{|c|c|c|c|c|c|c|}
    \caption{Optimal radii and corresponding potential energy for Lennard-Jones particles and Morse particles with shape parameter $\alpha = 60 / r_0,$ where $r_0$ is the distance at which the pair potential has its minimum. \label{tab:energies-radii}} \\
    \hline
    $N$ & $R_{LJ}^* / r_0$ & $U_{LJ}/N/\epsilon$ & $R_{LJ}^*/r_0$ (Voogd) & $U_{LJ}/N/\epsilon$ (Voogd) & $R_{M}^* / r_0$& $U_{M}^*/N/\epsilon$ \\
    \hline
    \endfirsthead
    \multicolumn{7}{c}%
    {\tablename\ \thetable\ -- \textit{Continued from previous page}} \\
    \hline
    $N$ & $R_{LJ}^* / r_0$ & $U_{LJ}/N/\epsilon$ & $R_{LJ}^*/r_0$ (Voogd) & $U_{LJ}/N/\epsilon$ (Voogd) & $R_{M}^* / r_0$& $U_{M}^*/N/\epsilon$ \\
    \hline
    \endhead
    \hline \multicolumn{7}{r}{\textit{Continued on next page}} \\
    \endfoot
    \hline
    \endlastfoot
    10 & 0.904877 & -2.386829 & 0.897777352534 & -2.391701447066 & 0.951257 & -2.099895 \\
    11 & 0.940335 & -2.546418 & 0.940905005832 & -2.546447320801 & 0.950446 & -2.268700 \\
    12 & 0.936682 & -2.795960 & 0.942373155294 & -2.799795573727 & 0.951435 & -2.498530 \\
    13 & 1.022253 & -2.448824 & 1.023669635577 & -2.448969977079 & 1.090487 & -1.966043 \\
    14 & 1.051020 & -2.532894 & 1.053553039689 & -2.533348412369 & 1.070727 & -1.999874 \\
    15 & 1.077747 & -2.604019 & 1.079511939730 & -2.604254157791 & 1.107120 & -1.996530 \\
    16 & 1.113089 & -2.629181 & 1.111359350373 & -2.629397082119 & 1.135076 & -1.997704 \\
    17 & 1.136519 & -2.729079 & 1.134715561189 & -2.729322104280 & 1.177153 & -2.108171 \\
    18 & 1.169474 & -2.720606 & 1.168431468281 & -2.720681589556 & 1.192575 & -2.143180 \\
    19 & 1.212959 & -2.692581 & 1.212803380000 & -2.692582368658 & 1.246011 & -2.162748 \\
    20 & 1.220995 & -2.812968 & 1.221102153780 & -2.812968910809 & 1.243276 & -2.238600 \\
    21 & 1.263170 & -2.759161 & 1.256842828104 & -2.761570772616 & 1.290012 & -2.033690 \\
    22 & 1.280587 & -2.813962 & 1.278935841127 & -2.814127147712 & 1.345578 & -2.309422 \\
    23 & 1.321960 & -2.798679 & 1.323636299622 & -2.798822415653 & 1.344108 & -2.390687 \\
    24 & 1.337533 & -2.916111 & 1.325942483975 & -2.923589586974 & 1.343885 & -2.499904 \\
    25 & 1.365106 & -2.789403 & 1.370215612837 & -2.790744681095 & 1.409161 & -2.143061 \\
    26 & 1.405526 & -2.831581 & 1.393253846649 & -2.838736266635 & 1.426258 & -2.064923 \\
    27 & 1.405019 & -2.915403 & 1.405226921913 & -2.915404947140 & 1.438392 & -2.174199 \\
    28 & 1.444699 & -2.834263 & 1.441037402745 & -2.834795918128 & 1.486233 & -2.109257 \\
    29 & 1.470268 & -2.836935 & 1.470703678855 & -2.836943974529 & 1.506661 & -2.157828 \\
    30 & 1.481021 & -2.907705 & 1.482942826361 & -2.907879985814 & 1.512497 & -2.264060 \\
    31 & 1.505503 & -2.914502 & 1.508252602796 & -2.914845518606 & 1.569113 & -2.171004 \\
    32 & 1.529593 & -2.973958 & 1.517799565208 & -2.980094374797 & 1.556748 & -2.156241 \\
    33 & 1.553309 & -2.851437 & 1.565483089900 & -2.857364512086 & 1.607324 & -2.090012 \\
    34 & 1.576668 & -2.879003 & 1.581507917153 & -2.879973993426 & 1.663424 & -2.215261 \\
    35 & 1.615217 & -2.892769 & 1.603568457619 & -2.897476940815 & 1.650354 & -2.254401 \\
    36 & 1.622380 & -2.926064 & 1.621869926831 & -2.926074652855 & 1.651993 & -2.304795 \\
    37 & 1.660724 & -2.902360 & 1.645689466050 & -2.907674813755 & 1.689376 & -2.167760 \\
    38 & 1.666836 & -2.970314 & 1.659845132456 & -2.972138629420 & 1.734758 & -2.155274 \\
    39 & 1.688627 & -2.913506 & 1.689160940137 & -2.913516422414 & 1.736032 & -2.163549 \\
    40 & 1.710142 & -2.935197 & 1.708230437867 & -2.935326506885 & 1.752469 & -2.195959 \\
    41 & 1.731381 & -2.929093 & 1.729343588049 & -2.929234832467 & 1.774234 & -2.146127 \\
    42 & 1.752371 & -2.957193 & 1.752712826656 & -2.957197169053 & 1.809763 & -2.265107 \\
    43 & 1.773111 & -2.985056 & 1.765631923177 & -2.986878602028 & 1.809023 & -2.237259 \\
    44 & 1.776194 & -3.038971 & 1.773591356738 & -3.039204045422 & 1.813941 & -2.271814 \\
    45 & 1.796266 & -2.988776 & 1.801077961986 & -2.989550533032 & 1.882977 & -2.388847 \\
    46 & 1.833924 & -2.967809 & 1.831567835991 & -2.967979946789 & 1.873818 & -2.265579 \\
    47 & 1.835750 & -2.988661 & 1.845624666183 & -2.991666695750 & 1.883939 & -2.442927 \\
    48 & 1.855172 & -3.043492 & 1.852527998434 & -3.043710551686 & 1.884393 & -2.498348 \\
    49 & 1.892777 & -2.959511 & 1.884936044669 & -2.961259366808 & 1.937731 & -2.177288 \\
    50 & 1.893427 & -2.983355 & 1.899868057549 & -2.984583094281 & 1.947870 & -2.204567 \\
    51 & 1.912270 & -2.991444 & 1.916517435225 & -2.991978366226 & 1.974846 & -2.198855 \\
    52 & 1.949857 & -2.984952 & 1.937974048925 & -2.988779822952 & 1.984468 & -2.187915 \\
    53 & 1.949402 & -2.970368 & 1.961895379520 & -2.974816531472 & 2.009351 & -2.195769 \\
    54 & 1.967710 & -3.007974 & 1.971909776092 & -3.008467736524 & 2.016416 & -2.273019 \\
    55 & 2.005315 & -2.972047 & 1.997949144047 & -2.973425131693 & 2.052942 & -2.222675 \\
    56 & 2.003818 & -3.005524 & 2.007994951024 & -3.005994640032 & 2.056417 & -2.243726 \\
    57 & 2.021627 & -2.993537 & 2.027327589626 & -2.994392848924 & 2.081772 & -2.257658 \\
    58 & 2.059277 & -2.991903 & 2.044677513303 & -2.997051140312 & 2.099946 & -2.252599 \\
    59 & 2.056791 & -3.009671 & 2.060451925427 & -3.010013049113 & 2.116954 & -2.302388 \\
    60 & 2.074146 & -3.032748 & 2.072914876984 & -3.032786493941 & 2.127555 & -2.282841 \\
    61 & 2.091367 & -3.004137 & 2.099984988857 & -3.005979486010 & 2.156723 & -2.251880 \\
    62 & 2.108437 & -3.021133 & 2.109459052546 & -3.021158756038 & 2.165891 & -2.216732 \\
    63 & 2.125373 & -3.022027 & 2.125802560541 & -3.022031670944 & 2.181170 & -2.228638 \\
    64 & 2.142175 & -3.020690 & 2.143018548376 & -3.020706178561 & 2.203763 & -2.233371 \\
    65 & 2.180011 & -3.014365 & 2.159849647419 & -3.023038735530 & 2.218712 & -2.252606 \\
    66 & 2.175388 & -3.034738 & 2.173766775635 & -3.034797798662 & 2.228147 & -2.242949 \\
    67 & 2.191807 & -3.033152 & 2.190095552387 & -3.033218086182 & 2.257012 & -2.223577 \\
    68 & 2.229750 & -3.009600 & 2.212957714507 & -3.015064415231 & 2.269333 & -2.223757 \\
    69 & 2.224280 & -3.034493 & 2.221952321872 & -3.034610208058 & 2.279338 & -2.228245 \\
    70 & 2.240334 & -3.042208 & 2.236128944654 & -3.042587333370 & 2.296906 & -2.255365 \\
    71 & 2.256281 & -3.047856 & 2.253609149146 & -3.048006785146 & 2.317771 & -2.299356 \\
    72 & 2.272121 & -3.056438 & 2.264321813954 & -3.057702042394 & 2.316996 & -2.325058 \\
    73 & 2.287837 & -3.019305 & 2.292383040075 & -3.019407304510 & 2.352293 & -2.226592 \\
    74 & 2.303454 & -3.030862 & 2.303497514639 & -3.030862374611 & 2.366227 & -2.242038 \\
    75 & 2.318974 & -3.044339 & 2.315406949106 & -3.044593521226 & 2.375216 & -2.212863 \\
    76 & 2.334377 & -3.043228 & 2.332264011995 & -3.043315920487 & 2.395653 & -2.245938 \\
    77 & 2.349683 & -3.051799 & 2.342682708578 & -3.052751366791 & 2.405489 & -2.261926 \\
    78 & 2.364891 & -3.067158 & 2.355651081550 & -3.068799176877 & 2.410416 & -2.237785 \\
    79 & 2.380009 & -3.046926 & 2.373368483194 & -3.047756931955 & 2.447245 & -2.252494 \\
    80 & 2.395021 & -3.053943 & 2.387441437503 & -3.055012900626 & 2.449508 & -1.902176 \\
    81 & 2.409943 & -3.040082 & 2.405496835696 & -3.040443111613 & 2.470801 & -2.195361 \\
    82 & 2.424777 & -3.036470 & 2.421820588972 & -3.036629161026 & 2.492485 & -2.259488 \\
    83 & 2.439512 & -3.039338 & 2.434795696871 & -3.039736810517 & 2.501109 & -2.328417 \\
    84 & 2.454168 & -3.051290 & 2.448182481981 & -3.051926932936 & 2.500183 & -2.345918 \\
    85 & 2.468734 & -3.041139 & 2.464409004970 & -3.041467105409 & 2.529841 & -2.209913 \\
    86 & 2.483211 & -3.044077 & 2.478604188594 & -3.044445037863 & 2.544674 & -2.224230 \\
    87 & 2.497608 & -3.048261 & 2.493077153951 & -3.048615801156 & 2.555677 & -2.204192 \\
    88 & 2.511916 & -3.058038 & 2.504241465416 & -3.059034645237 & 2.571579 & -2.265073 \\
    89 & 2.526152 & -3.053892 & 2.519344738182 & -3.054668510385 & 2.582359 & -2.252085 \\
    90 & 2.540300 & -3.049875 & 2.540613112184 & -3.049877137472 & 2.606983 & -2.251916 \\
    91 & 2.554376 & -3.052519 & 2.551677109091 & -3.052638760612 & 2.617603 & -2.210665 \\
    92 & 2.568372 & -3.067097 & 2.559448724569 & -3.068391387274 & 2.616534 & -2.281080 \\
    93 & 2.582297 & -3.061326 & 2.578325483646 & -3.061580134667 & 2.644918 & -2.250411 \\
    94 & 2.596141 & -3.069246 & 2.588610373792 & -3.070146514964 & 2.651315 & -2.127731 \\
    95 & 2.609915 & -3.070577 & 2.600111381806 & -3.072084399989 & 2.669293 & -2.297233 \\
    96 & 2.623617 & -3.078376 & 2.609596703568 & -3.081425625804 & 2.678059 & -2.287996 \\
    97 & 2.637247 & -3.074433 & 2.626721288869 & -3.076135067207 & 2.695931 & -2.298052 \\
    98 & 2.650807 & -3.087128 & 2.635565739021 & -3.090663894976 & 2.697864 & -2.282845 \\
    99 & 2.664295 & -3.066456 & 2.653858334618 & -3.068095087771 & 2.727576 & -2.205642 \\
    100 & 2.677712 & -3.072264 & 2.663546522523 & -3.075249310690 & 2.743986 & -2.232190 \\
\end{longtable}
\end{center}

\section{Cut-off radii for distance criterion}
In this section we present the cut-off radii we used for the nearest neighbour distance criterion. The distances $r^*$ are chosen to coincide with the minimum after the first peak in the pair distribution function. For those particle numbers $N$ where the first peak was split, we chose $r^*$ so that both split peaks are within $r^*.$ They are tabulated in Tab. \ref{tab:gr-cutoffs}.

\begin{center}
  \begin{longtable}{|c|c|c||c|c|c|}
    \caption{Cut-off radii used for the distance neighbour criterion for Lennard-Jones particles and Morse particles with shape parameter $\alpha = 60 / r_0,$ where $r_0$ is the distance at which the pair potential has its minimum. \label{tab:gr-cutoffs}} \\
    \hline
    $N$ & $r^*$ (LJ) & $r^*$ (Morse) & $N$ & $r^*$ (LJ) & $r^*$ (Morse) \\
    \hline
    \endfirsthead
    \multicolumn{6}{c}%
    {\tablename\ \thetable\ -- \textit{Continued from previous page}} \\
    \hline
    $N$ & $r^*$ (LJ) & $r^*$ (Morse) & $N$ & $r^*$ (LJ) & $r^*$ (Morse) \\
    \hline
    \endhead
    \hline \multicolumn{6}{r}{\textit{Continued on next page}} \\
    \endfoot
    \hline
    \endlastfoot
    10 & 1.1522287 & 1.1403504  & 11 & 1.1284720 & 1.1997439 \\ 
    12 & 1.0750181 & 1.0809568  & 13 & 1.1819259 & 1.0809568 \\ 
    14 & 1.3500679 & 1.1522287  & 15 & 1.3500679 & 1.0809568 \\ 
    16 & 1.1700467 & 1.1284720  & 17 & 1.0987748 & 1.0809568 \\ 
    18 & 1.0631388 & 1.0809568  & 19 & 1.1047144 & 1.0928361 \\ 
    20 & 1.0572001 & 1.0809568  & 21 & 1.1759863 & 1.0809568 \\ 
    22 & 1.0572001 & 1.0809568  & 23 & 1.2235006 & 1.0809568 \\ 
    24 & 1.2710158 & 1.0809568  & 25 & 1.1225324 & 1.1165928 \\ 
    26 & 1.1581683 & 1.1225324  & 27 & 1.1878647 & 1.0868964 \\ 
    28 & 1.2472582 & 1.0393821  & 29 & 1.1759863 & 1.0928361 \\ 
    30 & 1.1581683 & 1.1225324  & 31 & 1.3363481 & 1.1106540 \\ 
    32 & 1.3363481 & 1.2472582  & 33 & 1.1938043 & 1.1938043 \\ 
    34 & 1.1522287 & 1.0809568  & 35 & 1.1344107 & 1.1165928 \\ 
    36 & 1.0809568 & 1.1759863  & 37 & 1.1700467 & 1.0809568 \\ 
    38 & 1.1700467 & 1.0809568  & 39 & 1.1581683 & 1.1581683 \\ 
    40 & 1.1819259 & 1.1522287  & 41 & 1.1700467 & 1.1848953 \\ 
    42 & 1.1700467 & 1.0809568  & 43 & 1.2472582 & 1.1938043 \\ 
    44 & 1.3363481 & 1.2472582  & 45 & 1.1284720 & 1.2472582 \\ 
    46 & 1.1819259 & 1.1670773  & 47 & 1.0928361 & 1.2650762 \\ 
    48 & 1.3363481 & 1.2650762  & 49 & 1.2531978 & 1.2235006 \\ 
    50 & 1.1284711 & 1.2205312  & 51 & 1.1700467 & 1.2472582 \\ 
    52 & 1.1819259 & 1.1165928  & 53 & 1.1462900 & 1.2472582 \\ 
    54 & 1.1581683 & 1.1522260  & 55 & 1.1641080 & 1.1522260 \\ 
    56 & 1.1759863 & 1.1670773  & 57 & 1.1819259 & 1.1938043 \\ 
    58 & 1.1403504 & 1.2383492  & 59 & 1.1641080 & 1.2472582 \\ 
    60 & 1.2353799 & 1.2472582  & 61 & 1.2591366 & 1.2205312 \\ 
    62 & 1.1878647 & 1.1938043  & 63 & 1.2116223 & 1.2472582 \\ 
    64 & 1.2413186 & 1.1848953  & 65 & 1.2710158 & 1.2650762 \\ 
    66 & 1.0750181 & 1.2160768  & 67 & 1.2413186 & 1.1938043 \\ 
    68 & 1.2650762 & 1.2027133  & 69 & 1.2116223 & 1.1938043 \\ 
    70 & 1.2472582 & 1.2294402  & 71 & 1.2472582 & 1.1938043 \\ 
    72 & 1.2918031 & 1.2027133  & 73 & 1.1700467 & 1.2116223 \\ 
    74 & 1.1641080 & 1.1635137  & 75 & 1.1641080 & 1.1848953 \\ 
    76 & 1.2027133 & 1.2116223  & 77 & 1.2413186 & 1.2135822 \\ 
    78 & 1.0868964 & 1.1670773  & 79 & 1.1522287 & 1.1848953 \\ 
    80 & 1.1670773 & 1.2160768  & 81 & 1.2591366 & 1.2918031 \\ 
    82 & 1.1700467 & 1.2027133  & 83 & 1.1700467 & 1.2828942 \\ 
    84 & 1.1047144 & 1.2739852  & 85 & 1.1759863 & 1.3363481 \\ 
    86 & 1.2413186 & 1.2472582  & 87 & 1.1165928 & 1.2472582 \\ 
    88 & 1.1047144 & 1.2472582  & 89 & 1.2175619 & 1.1492593 \\ 
    90 & 1.1047144 & 1.2561672  & 91 & 1.2591366 & 1.2561672 \\ 
    92 & 1.1938043 & 1.2294402  & 93 & 1.1909445 & 1.2561672 \\ 
    94 & 1.1878647 & 1.2650762  & 95 & 1.1759863 & 1.2294402 \\ 
    96 & 1.2531978 & 1.1581683  & 97 & 1.2531978 & 1.2472582 \\ 
    98 & 1.1789441 & 1.2294402  & 99 & 1.2472582 & 1.2353790 \\ 
    100 & 1.2365407 & 1.2472582 &    &           & \\
\end{longtable}
\end{center}

\section{Free energies of packings}
This section contains detailed descriptions related to the identification of the packings considered in Section 4. To accurately identify the equilibrium packing, we average the particle positions over a short time window to average out fast thermal fluctuations. In particular, we average 250 frames 10 time steps apart, which corresponds to time intervals of $0.05 \tau_L$ apart, where $\tau_L$ is the Langevin damping time. This averaging is performed every 2500 time steps, which corresponds to intervals of $12.5 \tau_L.$

For each of these averaged snapshots, we determine the number of nearest neighbours of each particle according to the neighbour criterion with $r^* = 1.2918031r_0.$ For each of these defects, we determine if they are in a defect cluster, where a cluster is defined as all defects that are direct or indirect neighbours of each other. From that information, one can already deduce if the packing is icosahedral, $D_{5h},$ or $D_3,$ since an icosahedral packing contains twelve defect clusters of one defect each, the $D_{5h}$ contains five clusters of six defects and two of one defect, and the $D_3$ packing contains six clusters of one defect and three clusters of six particles.

With the aforementioned analysis, we can determine for each frame the packing, and from this we construct occurrence frequencies for each packing. Under the assumption that the simulations are ergodic, these can be converted into free energy differences, as explained in the main text.


\end{document}

%% file: paper_new4.bbl
\providecommand*{\mcitethebibliography}{\thebibliography}
\csname @ifundefined\endcsname{endmcitethebibliography}
{\let\endmcitethebibliography\endthebibliography}{}